\documentclass{aastex631}
\usepackage{natbib}
\usepackage{xcolor}

\usepackage{amsmath}

\graphicspath{{./}{figures/}}

\begin{document}
\title{Interstellar planetesimals: potential seeds for planet formation?}

\author{Amaya Moro-Mart\'{\i}n}
\affiliation{Space Telescope Science Institute, 3700 San Martin Dr., Baltimore, MD 21218}
\affiliation{Physics and Astronomy, Center for Astrophysical Sciences, Johns Hopkins University, Baltimore, MD 21218}

\author{Colin Norman}
\affiliation{Space Telescope Science Institute, 3700 San Martin Dr., Baltimore, MD 21218}
\affiliation{Physics and Astronomy, Center for Astrophysical Sciences, Johns Hopkins University, Baltimore, MD 21218}

\begin{abstract}
We investigate the trapping of interstellar objects during the early stages of star and planet formation. Our results show a very wide range of possible values that will be narrowed down as the population of interstellar objects becomes better characterized. When assuming a background number density of 2$\cdot$10$^{15}$ pc$^{-3}$ (based on 1I/'Oumuamua detection), a velocity dispersion of 30 km/s and an equilibrium size distribution, the number of interstellar objects captured by a molecular cloud and expected to be incorporated to each protoplanetary disk during its formation is
O(10$^{9}$) (50 cm--5 m), 
O(10$^{5}$)  (5 m--50 m), 
O(10$^{2}$)  (50 m--500 m), 
O(10$^{-2}$)  (500 m--5 km). After the disk formed, the number of interstellar objects it could capture from the ISM during its lifetime is  
6$\cdot$10$^{11}$ (50 cm--5 m), 
2$\cdot$10$^{8}$ (5 m--50 m), 
6$\cdot$10$^{4}$ (50 m--500 m), 
20 (500 m--5 km); in an open cluster where 1\% of stars have undergone planet formation, these values increase by a factor of O(10$^{2}$--10$^{3}$). These trapped interstellar objects might be large enough to rapidly grow into larger planetesimals via the direct accretion of the sub-cm sized dust grains in the protoplanetary disk before they drift in due to gas drag, helping overcome the meter-size barrier, acting as "seeds" for planet formation. They should be considered in future star and planet formation models and in the potential spread of biological material across the Galaxy. 
\end{abstract}


\section{Introduction}
\label{Introduction}

Stars form from the collapse of dense regions of molecular clouds. A natural byproduct of this process is the formation of protoplanetary disks, where, according to the leading theory of planet formation, planet formation takes place (Hayashi \citeyear{1981PThPS..70...35H}; Shu et al. \citeyear{1987ARA&A..25...23S}; Hartmann \citeyear{Hartmann2008}). The dust particles in these disks, approximately 0.1 $\mu$m in size, are strongly coupled to the gas, resulting in small relative velocities and collisional energies that, together with "sticky" microphysical processes (like van der Waals and electromagnetic forces), result in their efficient collisional growth into cm-sized bodies (Weidenschilling \citeyear{1977MNRAS.180...57W}, \citeyear{1980Icar...44..172W}; Birnstiel et al. \citeyear{2011A&A...525A..11B}); these pebbles eventually grow into km-sized planetesimals and the processes by which these can grow into planetary embryos and planets, via collisions and gravitational interactions, is fairly well understood(Armitage \citeyear{2010apf..book.....A}, and references therein).  However, the intermediate stage by which cm-sized pebbles grown into km-sized planetesimals poses several challenges, known as the meter-sized barrier. As the cm-sized particles grow and become less coupled to the gas, their relative velocities and collisional energies increase, resulting in collisions that, rather than leading to efficient growth, lead to inefficient sticking, bouncing, or fragmentation (Zsom et al. \citeyear{2010A&A...513A..57Z}); in addition, because the particles are still coupled to the gas, they experience gas drag that results in short inward drift timescales. Weidenschilling (\citeyear{1977MNRAS.180...57W}) found that a 1 m-sized particles with a density of 3 g$\cdot$cm$^{-3}$ located at a radial distance of 5 AU in a solar nebula has a drift lifetime of O(10$^{2})$ years. This short drift timescales limits significantly their lifetime in the disk  and, consequently, their opportunity to grow to sizes unaffected by gas drag (Birnstiel et al. \citeyear{2012A&A...539A.148B}). In addition, the aeolian erosion that small planetesimals experience in the presence of gas can constitute a significant barrier for the growth of metre-size objects in protoplanetary discs (Rozner et al. \citeyear{2020MNRAS.496.4827R}) and could contribute to a change of the planetesimal mass function and particle concentration at a unique Stokes number (Grishin et al. \citeyear{2020ApJ...898L..13G}).

Several mechanisms have been proposed to alleviate the meter-sized barrier: the effect of the gravitational collapse of the solid component in the disk (Youdin \& Su \citeyear{2002ApJ...580..494Y}; Youdin \& Goodman \citeyear {2005ApJ...620..459Y}) and the formation of over-dense filaments in the disk fragmenting gravitationally into clumps of small particles that collapse and form planetesimals (Johansen et al. \citeyear {2007Natur.448.1022J}); the effect of the velocity distribution of the dust particles that allows for the existence of low-velocity collisions that favor the growth of a small fraction of particles into larger bodies (Windmark \citeyear{2012A&A...544L..16W}); the presence of pressure maxima that can trap dust grains (Johansen et al. \citeyear{2004A&A...417..361J}); the effect of the low porosity (Okuzumi et al. \citeyear{2009ApJ...707.1247O}) and the electric charging of dust aggregates (Okuzumi \citeyear{2009ApJ...698.1122O}) to favor sticking over fragmentation;  and collisional growth between particles of very different masses colliding at high velocity (Booth et al. \citeyear{2018MNRAS.475..167B}). 

The detections of 1I/'Oumuamua and 2I/Borisov, the first interstellar interlopers, have pointed out to the existence of an additional factor that could help overcome the challenges that the meter-sized barrier imposes on planetesimal and planet formation models: the seeding of the disks with planetesimals captured or incorporated from the interstellar medium. Grishin et al. (\citeyear{2019MNRAS.487.3324G}) studied the scenario in which interstellar objects are captured by the protoplanetary disk due to gas drag, while Pfalzner \& Bannister (\citeyear{2019ApJ...874L..34P}) studied how, during the cloud-formation process, interstellar objects could be incorporated to the molecular cloud in the same way as the interstellar gas and dust, and from there could be incorporated to the the star-forming disks. 

In this work, we study the trapping of interstellar objects by gas drag in several environments characteristic of the star- and planet-formation processes, including a molecular cloud, a stellar kernel and a protoplanetary disk. We improve upon Pfalzner \& Bannister (\citeyear{2019ApJ...874L..34P}) work by considering the effect of gas drag in the trapping of the interstellar objects. In Section \ref{ISO capture ISM} we describe the trapping environments and characteristic incoming velocity; in Section \ref{capturerates} we calculate the probability to capture interstellar objects as a function of the environmental parameters and the properties of the objects, taking into account the distribution of their incoming velocities. In Section \ref{totalnumber}, we use this capture probability to derive an expression for the total number of captured planetesimals, that we improve in Section \ref{improvedcapturerate} taking into account the trajectory and orbital ellipticity of the particle as it traverses the environment. In Section \ref{molecularcloud} we estimate the total number of interstellar objects that could be captured in the molecular cloud environment and incorporated to each stellar system, and in Section \ref{kernelmmsn} we do the same for the pre-stellar kernel and protoplanetary disk environments. In Section \ref{ISO capture cluster}, we consider the case of an open cluster and in Section \ref{discussion} we discuss the results. 

In these calculations, we adopt two different assumptions regarding the origin of the interstellar objects and their expected background density: (1) a background density of interstellar objects of $N_{\rm R\geqslant R_{\rm Ou}} = 2\cdot10^{15} {\rm pc}^{-3}$, with $R_{\rm Ou} \sim$ 80 m, inferred from 1I/'Oumuamua's detection (Do et al. \citeyear{2018ApJ...855L..10D}); (2) a lower value of m$_{\rm total}$ = 9.5$\cdot$10$^{27}$ g$\cdot$pc$^{-3}$ corresponding to the background density that would be expected from the ejection of planetesimals from protoplanetary disks (Moro-Mart{\'{\i}}n \citeyear {2018ApJ...866..131M}). 

\section{Trapping environments and incoming velocities}
\label{ISO capture ISM}
To explore the effect of the gas drag, we start by considering a uniform density, spherically symmetric molecular cloud and proto-stellar core kernel, the latter similar to those recently resolved by ALMA (Caselli et al. \citeyear{2019ApJ...874...89C}). Table 1 lists their physical properties, namely their number density, $n_{\rm env}$, Temperature, $T_{\rm env}$, cloud radius $R_{\rm env}$ and estimated lifetime, $\tau_{\rm env}$. We also consider a more dense and compact environment with gas densities and temperatures characteristic of the mid-plane of a minimum mass solar nebula (MMSN) at 30 au (Hayashi \citeyear{1981PThPS..70...35H}) and a protoplanetary disk (Chiang \& Goldreich \citeyear{1997ApJ...490..368C}). For the disk case we  assume a uniform density and that the swept up column is that along the scale height. A detailed study that
takes into account the disk structure is given by Grishin et al. (\citeyear{2019MNRAS.487.3324G}).

The mean velocity dispersion of molecular clouds in the Galaxy is in the range 1--10 km/s (see e.g. Fig. 5 in Dobbs et al. \citeyear{2019MNRAS.485.4997D}), while the velocity distribution of young stars in the galaxy is characterized by a dispersion $\sigma_R \sim$ 30 km/s (Mackereth et al. \citeyear{2019MNRAS.489..176M}). In the solar neighborhood, the rotational frequency of the Galaxy is similar to that of the spiral arms (Gerhard \citeyear{2011MSAIS..18..185G}; Nguyen et al. \citeyear{2018MNRAS.475...27N}) since the solar neighborhood is near corotation. This means that, even if we were to assume that star and planet formation preferentially takes place in the spiral arms (which is strictly only true for high-mass stars), we can ignore the streaming velocity of the stars.

Because of corotation, and also because low-mass star formation happens in molecular clouds in-between arms, let us assume that there is no streaming velocity and that $u$, the characteristic incoming velocity of interstellar objects as they enter the environment under consideration, is in the 1--30 km/s range, given by the velocity dispersion mentioned earlier. The situation would be different closer to the Galactic center because at $\lesssim$ 4 kpc, the the streaming velocity itself would be $\sim$ 100 km/s (Nguyen et al. \citeyear{2018MNRAS.475...27N}) and would need to be taken into account.

\startlongtable 
\begin{deluxetable}{llllll}
\tablenum{1}
\tablecaption{Environments under consideration.}
\tablewidth{0pc}
\tablehead{
\colhead{Environment} &
\colhead{$n_{\rm env}$} &
\colhead{$T_{\rm env}$} &
\colhead{$R_{\rm env}$} &
\colhead{$\tau_{\rm env}$} & 
\colhead{$u$} 
\\
\colhead{} &
\colhead{(cm$^{-3}$)} &
\colhead{(K)} &
\colhead{} &
\colhead{(Myr)} &
\colhead{(km/s)}
}
\startdata
Molecular cloud 	    			& 10$^{2}$		        	& 10		& 15 pc		& 30 					& 1--30	\\
Pre-stellar kernel				& 10$^{6}$		        	& 6.5		& 1500 AU	& 1\tablenotemark{a} 	& 1--30	\\
MMSN-30au\tablenotemark{b}		& 8$\cdot$10$^{10}$		& 50		& 30 AU	    	& 1\tablenotemark{a} 	& 1--30	\\
DISK-30au\tablenotemark{c}		& 5$\cdot$10$^{9}$		& 50		& 30 AU	    	& 1\tablenotemark{a} 	& 1--30	\\
\enddata
\tablenotetext{a}{The pre-stellar core kernel is expected to be short-lived, with a lifetime of the order of its free-fall timescale (Andr{\'e} et al. \citeyear{2014prpl.conf...27A}). The disk lifetime is 1--10 Myr; here we assumed 1 Myr.}
\tablenotetext{b}{Gas density and temperature characteristic of the mid-plane of a minimum mass solar nebula at 30 au (Hayashi\citeyear{1981PThPS..70...35H}).}
\tablenotetext{c}{Gas density and temperature characteristic of the mid-plane of a protoplanetary disk at 30 au, based on Chiang \& Goldreich (\citeyear{1997ApJ...490..368C}), following Grishin et al. (\citeyear{2019MNRAS.487.3324G}). See Section \ref{kernelmmsn} for discussion; this disk is thicker than the one considered for the MMSN case.}
\end{deluxetable}

\section{Gas drag and Capture Rates}
\label{capturerates}

\begin{figure}
\begin{center}
\includegraphics[width=9cm]{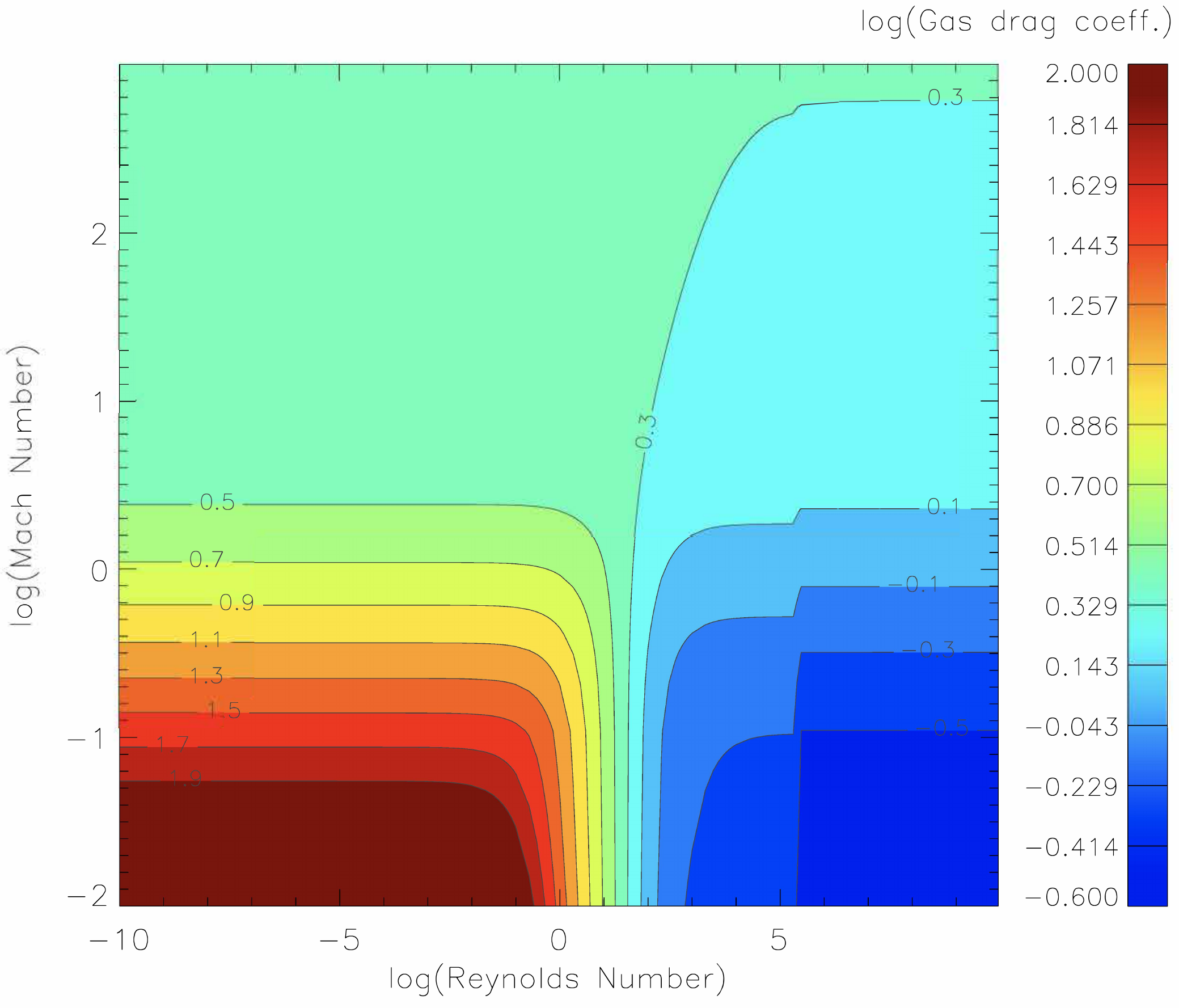}
\end{center}
\caption{Gas drag coefficient as a function of the Reynolds number and the Mach number, calculated using Equation \eqref{Cd}.}
\label{FCd}
\end{figure}

The acceleration that an interstellar object will experience due to gas drag is 
\begin{equation}
a_{\rm D} = {1\over2 m_{\rm p} }C_{\rm D}\pi r_{\rm p}^{2} \rho_{env} u^{2} = {3 C_{\rm D} \rho_{env} u^{2} \over 8 \rho_{\rm p} r_{\rm p}}, 
\label{aD}
\end{equation}
where $\rho_{\rm env}$ is the mass density of the environment ($\rho_{\rm env} = n_{\rm env} m_{0} $, with $m_{0}$ = 3.34$\cdot$10$^{-24}$ g for for H$_2$), and $m_{\rm p}$, $r_{\rm p}$, $\rho_{\rm p}$, and $u$ are the mass, radius, density, and incoming velocity of the interstellar object, respectively. $C_{D}$ is the gas drag coefficient, a function of the Mach number, $M = u/c$, and the Reynolds numbers, $R_{\rm e}$, that can be approximated by 
\begin{equation}
\begin{split}
C_{\rm D} \simeq \left(\left({24 \over R_{\rm e}} + {40 \over 10+R_{\rm e}}\right)^{-1} + {0.23~M}\right)^{-1} 
+ {(2-w) M  \over (1.6+M)} + w,
\end{split}
\label{Cd}
\end{equation}
where  $w$ = 0.4, if $R_{e} < 2\cdot$10$^{5}$, and $w$ = 0.2, if $R_{e} > 2\cdot$10$^{5}$  (Adachi et al. \citeyear{1976PThPh..56.1756A}; Tanigawa et al. \citeyear{2014ApJ...784..109T}). Figure \ref{FCd} shows the gas drafg coefficient as a function of the Reynolds number and the Mach number, calculated with Equation \ref{Cd}. In Appendix A and Table A1 we calculate, for each environment and for a given interstellar object incoming velocity, the Mach number, the Reynolds number, and the Stokes number. The latter is the ratio of the stopping time to a dynamical time; interstellar objects with $Stk <<$ 1 are expected to be trapped by gas drag, while those with $Stk >>$ 1 are expected to escape (see discussion in Appendix A).

To calculate the capture rates we start by assuming, as in Grishin et al. (\citeyear{2019MNRAS.487.3324G}), that the relative velocity $u$ is a constant through the passage so that the energy dissipated by the particle as it crosses the environment is given by
\begin{equation}
\Delta E  = -{1\over2} \int {C_{\rm D}(u)\pi r_{\rm p}^{2} \rho_{env} u^{2}} {\rm d}x 
\approx -{C_{\rm D}(u)\pi r_p^2 \over 2}\Sigma_{env}u^2, 
\label{deltaE}
\end{equation}
where $u$ is the relative velocity and we let $u^2 = u_0^2 + u_{\rm esc}^2$ (see Grishin et al. \citeyear{2019MNRAS.487.3324G}, Equation (4)). Here, $u_{\rm esc} = \sqrt{2 G M \over R_{\rm env}}$; for the molecular cloud we assume that $M$ is the mass of the cloud but for the pre-stellar core kernel and the protoplanetary disk we will assume a solar mass at the center. Capture takes place when the energy loss is larger than the particle's kinetic energy at infinity, $\Delta E > E_0 = {1\over2} m_p u_0^2$. This leads to the capture condition, 
\begin{equation}
\begin{split}
{3 \over 4}{C_{\rm D}(u_0) \Sigma_{\rm env}  \over \rho_p r_p } (1+ {u_{\rm esc}^2 \over u_0^2}) > 1, \\
\label{capcondgen}
\end{split}
\end{equation}
which is a condition on the particle's size. 

It is a significant assumption that the velocity-dependent $C_D$ can be taken outside the energy integral and evaluated at a specific velocity. In Appendix B we derive the energy loss condition when one does not make this assumption; we discuss two limiting cases: a circular orbit grazing the cloud and a harmonic oscillator orbit passing through the center of the cloud. It is also a significant assumption that the relative velocity $u$ is a constant through the passage. In Section \ref{improvedcapturerate} we improve upon this assumption by deriving the capture rate taking into account the change in the relative velocity as the particle traverses the environment.

\subsection{Geometric regime}
In the geometric regime, ${u_{\rm esc}^2 \over u_0^2} << 1$, the trajectory of the particle follows a straight line and the column it sweps up  is $\Sigma_{\rm env} = 2 \rho_{\rm env} \sqrt{R_{\rm env}^2-b^2}$, where $b$ is the impact parameter. Following Equation \eqref{capcondgen}, in this case the condition for capture is 
\begin{equation}
{3 \over 2}{C_{\rm D}(u_0) \over \rho_p r_p} \rho_{\rm env} \sqrt{R_{\rm env}^2-b^2}  > 1,  
\label{capcondgeo}
\end{equation}
leading to a capture condition for the impact parameter of 
\begin{equation}
b(r_p) < b_c(r_p) = \sqrt {R_{\rm env}^2 - \left({2 \rho_p r_p \over 3 C_{\rm D}(u_0) \rho_{\rm env}}\right)^2}.
\label{capcondgeob}
\end{equation}

Following Grishin et al. (\citeyear{2019MNRAS.487.3324G}), the capture probability is 
\begin{equation}
f_c(r_p) =  \left({b_c(r_p) \over b_{\rm max}}\right)^2 
= {1\over b_{\rm max}^2} \left(R_{\rm env}^2 - \left({2 \rho_p r_p \over 3 C_{\rm D}(u_0) \rho_{\rm env}}\right)^2\right),  
\end{equation}
where $b_{\rm max}$ is the maximal impact parameter for an effective close encounter; in the geometrical regime, $b_{\rm max} = R_{\rm env}$ and the capture probability is independent of the particle's velocity (except for the $u$ dependence in the drag coefficient). 

\subsection{Gravitational focusing regime}
In the gravitational focusing regime, ${u_{\rm esc}^2 \over u_0^2} >> 1$,  for simplicity we will assume the particle enters in a  parabolic trajectory. In the grazing limiting case, the particle is trapped in a circular orbit of radius $R_{\rm env}$, with swept up column of $2 \pi R_{\rm env}$.  In the radial orbit limiting case, the particle traverses the cloud and then reverses its course, with a swept up column density of 4$R_{\rm env}$. In general, the orbit is elliptical and is a harmonic oscillator in $r^2$ (see Harter Soft Web Resources for Physics, Chapter 5). Therefore, we parameterize the swept up column density as $\Sigma_{\rm env} = 2 \xi \pi R_{\rm env} \rho_{\rm env}$, where $\xi$ is O(1). The capture condition based on energy loss is then 
\begin{equation}
\begin{split}
{3 \over 2}{C_{\rm D} \pi \xi R_{\rm env} \rho_{\rm env} \over \rho_p r_p } {u_{\rm esc}^2 \over u_0^2} > 1,\\
u_0^2 < u_{\rm 0,c}^2 = {3 \over 2}{C_D \pi \xi R_{\rm env} \rho_{\rm env} \over \rho_p r_p } u_{\rm esc}^2.
\label{capcondgf}
\end{split}
\end{equation}
Henceforth, we will evaluate the above expression at $\xi$ = 1. 

From angular momentum conservation, we have that 
\begin{equation}
b^2 u_0^2 =  a^2 u_{\rm esc}^2(a) = {2 G M_{\rm env} a},  
\label{b}
\end{equation}
giving a maximum limit on the angular momentum of the trapped object. 
To be trapped we have that $a < R_{\rm env}$, leading to ${b^2 u_0^2 \over 2 G M_{\rm env}} < R_{\rm env}$. The capture condition on the impact parameter is
\begin{equation}
\begin{split}
b(r_p)^2 < b_c(r)^2 = {2 G M_{\rm env} R_{\rm env} \over u_0^2} = {u_{\rm esc}^2 \over u_0^2} R_{\rm env}^2.
\end{split}
\label{capcondgeob}
\end{equation}

Following Grishin et al. (\citeyear{2019MNRAS.487.3324G}), the weighted capture probability is given by
\begin{equation}
f_c =  \int\int_{D,c} {2b \over b_{\rm max}^2} {u^3 \over 2 \sigma^4 } e^{-{u^2 \over 2 \sigma^2}} dbdu,   
\end{equation}
(see their Equation 16). The capture probability is weighted with an additional $u$ factor to take into account that faster planetesimals have higher encounter rates than slower ones. Here, $\sigma$ is the velocity dispersion of the velocity distribution. $b_{\rm max}$ is the maximal impact parameter for an effective close encounter, that in the gravitational focusing regime is $b_{\rm max} = R_{\rm env}{u_{\rm esc}\over u_0}$, where in this calculation we are now assuming ${u_{\rm esc}\over u_0} > $1. $D_{c}$ is the integration domain that results in capture, i.e., the integral over $u$ goes up to $u_{\rm 0,c}$ (Equation \eqref{capcondgf}). The domain for impact parameter is determined by the angular momentum condition, and the domain for the velocity $u_0$ is determined by the energy condition in the $b-u_0$ plane. In this case, the domain is clear because the spherical cloud is finite, unlike the infinite disk in  Grishin et al. (\citeyear{2019MNRAS.487.3324G}). The limiting value of $b$ is given by Equation \eqref{capcondgeob}, and the limiting value of $u_0$ is given by Equation \eqref{capcondgf}. 

We then have,
\begin{equation}
\begin{split}
f_c (r_p, a) = {1 \over b_{\rm max}^2} \int^{u_{\rm 0,c}}_{0} du {u^3 \over 2 \sigma^4 } e^{-{u^2 \over 2 \sigma^2}} \int_0^{b_c} 2b db 
=  {1 \over b_{\rm max}^2} \int^{u_{\rm 0,c}}_{0} du {u^3 \over 2 \sigma^4 } e^{-{u^2 \over 2 \sigma^2}}R_{\rm env}^2 {u_{\rm esc}^2(a) \over u^2}
= {R_{\rm env}^2 \over b_{\rm max}^2}{u_{\rm esc}^2 \over 2 \sigma^2}  \int^{{u_{\rm 0,c}}^2 \over 2 \sigma^2}_{0} dt  e^{-t}\\
\end{split}
\end{equation}
\begin{equation}
f_c (r_p) =  {R_{\rm env}^2 \over b_{\rm max}^2}{u_{\rm esc}^2 \over 2 \sigma^2} \left[1-e^{-{{u_{\rm esc}^2} \over \sigma^2}{3 \over 4}{C_D\pi R_{\rm env}\rho_{\rm env} \over \rho_p r_p }}\right].
\label{fc}
\end{equation}
If we define an ensemble average Safronov number ${\bar\Theta_s} = {u_{esc}^2\over \sigma^2}$, then 
\begin{equation}
f_c (r_p) =  {R_{\rm env}^2 \over 2b_{\rm max}^2}{\bar\Theta_s}\left[1-e^{-{\bar\Theta_s} {3 \over 4}{C_D\pi R_{\rm env}\rho_{\rm env} \over \rho_p r_p }}\right].
\label{fctheta}
\end{equation}

For small arguments of the exponential in Equation \eqref{fctheta}, we can simplify the capture rate to
\begin{equation}
\begin{split}
f_c (r_p) =  {3 \over 8} \pi C_D {R_{\rm env}^2 \over b_{\rm max}^2}{\bar\Theta_s}^2 {R_{\rm env}\rho_{\rm env} \over \rho_p r_p }
=  {9 \over 32} {C_D \over b_{\rm max}^2}{\bar\Theta_s}^2 {M_{\rm env}\over \rho_p r_p}.
\label{fcsmall}
\end{split}
\end{equation}
Note that the capture rate is for sufficient dissipation in one orbital period at $R_{\rm env}$. In the case of a constant density cloud, that orbital period is the same across the cloud. Inner orbits have less dissipation because the circumference is smaller for one orbital period. 

It is important to note that in the calculations above, in the gravitational focusing case $C_D$ is evaluated at $u_{\rm esc}$ while in the geometric case $C_D$ is evaluated at $u_0$. This makes it difficult to generate a formula interpolating neatly between the two cases. 

\section{Total number of captured planetesimals}
\label{totalnumber}
The total number of interstellar objects that can be trapped will be given by $N(r_p) = f_c(r_p) \cdot N_{\rm enter}(r_p)$, where $f_c(r_p)$ is the weighted capture probability calculated above, so that  
\begin{equation}
N(r_p) =  {R_{\rm env}^2 \over b_{\rm max}^2}{u_{\rm esc}^2 \over 2 \sigma^2} \left[1-e^{-{{u_{\rm esc}^2} \over \sigma^2}{3 \over 4}{C_D\pi R_{\rm env}\rho_{\rm env} \over \rho_p r_p }}\right] {N_{\rm enter}(r_p)}.
\label{Nrptclarge}
\end{equation}
$N_{\rm enter}(r_p)$ is the number of interstellar objects of radius $r_p$ that enter the cross-section $\pi b_{\rm max}^2$, $N_{\rm enter}(r_p) = n_{\rm ISO}(r_p) \pi b_{\rm max}^2\langle u \rangle\tau_{\rm env}$. Here, $b_{\rm max}$ is the maximal impact parameter for an effective close encounter, $n_{\rm ISO}(r_p)$ is the number density of interstellar objects of radius $r_p$, $\tau_{\rm env}$ is the lifetime of the cloud, i.e. the timescale over which encounters can occur, and $\langle u \rangle = \sqrt{8 \over \pi} \sigma$ is the mean velocity of the incoming interstellar objects, where $\sigma$ is the velocity dispersion. 
From Equation \eqref{Nrptclarge}, we then have that, 
\begin{equation}
\begin{split}
N(r_p) =  {\sqrt {2 \pi}}n_{\rm ISO}(r_p) \tau_{\rm env} R_{\rm env}^2 {u_{\rm esc}^2 \over \sigma} 
 \left[1-e^{-{{u_{\rm esc}^2} \over \sigma^2}{3 \over 4}{C_D\pi R_{\rm env}\rho_{\rm env} \over \rho_p r_p }}\right],
\label{NrptclargeISO}
\end{split}
\end{equation}
where for small arguments of the exponential we can simplify the total number of captured planetesimals to
\begin{equation}
\begin{split}
N(r_p) = {3 \pi \over 4} {\sqrt {2 \pi}}{C_D} n_{\rm ISO}(r_p) \tau_{\rm env} R_{\rm env}^2 u_{\rm esc} 
 \left({u_{\rm esc} \over \sigma}\right)^3{R_{\rm env}\rho_{\rm env} \over \rho_p r_p }.
\label{NrpISOfinal}
\end{split}
\end{equation}

To get to this expression for $N(r_p)$ we assumed in the derivation that the relative velocity $u$ is a constant through the passage. In the following section we improve upon this assumption by deriving the capture rate and the total number of captured planetesimals taking into account the change in the relative velocity as the particle traverses the environment.

\section{Improved calculation of the  total number of captured planetesimals taking into account the trajectory and orbital ellipticity of the particle}
\label{improvedcapturerate}

A uniform density environment, as the one that we are considering here, has a harmonic oscillator gravitational potential and in this potential the orbit of the particle is elliptical and is a harmonic oscillator in $r^2$ (see Harter Soft Web Resources for Physics, Chapter 5), $r^2 = {E \over m_p \Omega^2} (1 + e cos(2 \Omega t))$, where $\Omega^2 = {4\over 3}\pi G\rho_{\rm env}$ and the eccentricity is $e = \left(1-{\Omega^2 L^2 \over E^2}\right)^{1/2}$. The energy is $E = {1\over 2} m_p u^2 + {1\over 2} m_p \Omega^2 r^2$ and from this expression we can solve for the velocity, $u^2 = 2{E\over m_p} - \Omega^2 r^2 = {E \over m_p} (1 - e cos (2 \Omega t))$. This allows to calculate the energy dissipated by the particle as it crosses the environment, taking into account the change in $u$, 
\begin{equation}
\Delta E  = -{1\over2} \int {C_{\rm D}(u)\pi r_{\rm p}^{2} \rho_{env} u^{2}} u {\rm d}t 
= -{1\over2} {C_{\rm D}(u_{\rm esc})\pi r_{\rm p}^{2} \rho_{env}  \int_0^{T/2} u^{3}}  {\rm d}t,
\label{deltaEu}
\end{equation}
where we integrate over half a period, with $T = {2\pi \over \Omega}$, and we approximate C$_D$ by evaluating it  at $u_{\rm esc}$ so it can be taken outside the integral. Using the expression for $u^2$ derived above, and taking into account that $\left({E\over m_p}\right)^{1/2} \approx u_{\rm esc}$ and  $u_{\rm esc}\cdot{T\over 2}$ is half the perimeter of the elliptical orbit, $u_{\rm esc}\cdot{T\over 2} \approx 2 a E(e)$, we have
\begin{equation}
\begin{split}
\int_0^{T/2}  u^3 {\rm d}t = \left({E \over m_p}\right)^{3/2} \int_0^{T/2}  (1 - e {\rm cos} (2 \Omega t))^{3/2} {\rm d}t 
= \left({E \over m_p}\right)^{3/2} {1\over 2 \Omega}\int_0^{2\pi}  (1 - e {\rm cos} \theta)^{3/2} {\rm d}\theta\\
= \left({E \over m_p}\right)\left({E \over m_p}\right)^{1/2} {T\over 2}{1 \over 2 \pi}\int_0^{2\pi}  (1 - e {\rm cos} \theta)^{3/2} {\rm d}\theta 
= \left({E \over m_p}\right)u_{\rm esc}{T\over 2}f_3(e) 
= \left({E \over m_p}\right)2aE(e)f_3(e) 
\label{int}
\end{split}
\end{equation}
where $\theta = 2 \Omega t$, E(e) is the complete elliptic integral of the second kind, and $f_3(e) = {1 \over 2 \pi}\int_0^{2\pi}  (1 - e {\rm cos} \theta)^{3/2} {\rm d}\theta$. Hereafter we will use $\Psi(e) \equiv E(e)\cdot f_3(e)$ that for $e$ ranging from 0 to 1 is $\Psi(e)$ = 1.20--1.56. In the calculations in Sections \ref{molecularcloud} and \ref{kernelmmsn} we use $\Psi(e)$ = 1.20. The energy dissipated by the particle as it crosses the environment is then 

\begin{equation}
\Delta E  = - C_{\rm D}(u_{\rm esc})\pi r_{\rm p}^{2} \rho_{\rm env} a\Psi(e) \left({E \over m_p}\right).
\end{equation}
Assuming as before that the relative velocity is $u^2 = u_0^2 + u_{\rm esc}^2$, where $u_{\rm esc} = \sqrt{2 G M_{\rm env} \over R_{\rm env}}$ (see Grishin et al. \citeyear{2019MNRAS.487.3324G}, Equation (4)), we have that  ${E \over m_p} = {1\over 2} \left(u_0^2 + u_{\rm esc}^2\right)$ so that 

\begin{equation}
\Delta E  =  - C_{\rm D}(u_{\rm esc})\pi r_{\rm p}^{2} \rho_{\rm env} a\Psi(e) {1\over2}\left(u_0^2 + u_{\rm esc}^2\right). 
\end{equation}

Capture takes place when the energy loss is larger than the particle's kinetic energy at infinity, $\Delta E > E_0 = {1\over2} m_p u_0^2$. This leads to the capture condition, 
\begin{equation}
\begin{split}
{3 \over 4}{C_{\rm D}(u_{\rm esc})a \rho_{\rm env}  \Psi(e) \over \rho_p r_p }\left(1+ {u_{\rm esc}^2 \over u_0^2}\right) > 1, \\
\label{capcondgenelli}
\end{split}
\end{equation}
which is a condition on the particle's size that is similar to that in Equation \eqref{capcondgen} when $a \sim R_{\rm env}$. The capture condition in the gravitational focussing regime, equivalent to Equation \eqref{capcondgf}, is then 
\begin{equation}
\begin{split}
u_{0}^2 < u_{0,c}^2 = {3 \over 4}{C_{\rm D}(u_{\rm esc})a \rho_{\rm env}  \Psi(e) \over \rho_p r_p }u_{\rm esc}^2. \\
\label{capcondgenelli}
\end{split}
\end{equation}
This trapping condition can only be satisfied when ${u_{\rm esc}^2 \over u_0^2} >> 1$, i.e.  low velocity particles with high Safronov numbers. Our average over the incoming interstellar object distribution takes into account this stringent trapping condition. Note that the average Safronov numbers that appear are the result of the gravitational focussing and the phase space factors at the low end of the velocity space distribution of the interstellar objects. 

Their weighted capture probability, equivalent to Equations \eqref{fctheta} and \eqref{fcsmall}, is
\begin{equation}
\begin{split}
f_c (r_p) =  {R_{\rm env}^2 \over 2b_{\rm max}^2}{\bar\Theta_s}\left[1-e^{-{\bar\Theta_s} {3 \over 8}{C_D R_{\rm env}\rho_{\rm env}  \Psi(e) \over \rho_p r_p }}\right]
\sim {9 \over 64 \pi} {C_D \over b_{\rm max}^2}{\bar\Theta_s}^2 {M_{\rm env}\Psi(e)\over \rho_p r_p}.
\end{split}
\label{fcthetaelli}
\end{equation}
The total number of interstellar objects that can be trapped, equivalent to Equations \eqref{Nrptclarge} and \eqref{NrptclargeISO}, is 
\begin{equation}
\begin{split}
N(r_p) =  {R_{\rm env}^2 \over b_{\rm max}^2}{u_{\rm esc}^2 \over 2 \sigma^2} \left[1-e^{-{{u_{\rm esc}^2} \over \sigma^2}{3 \over 8}{C_D R_{\rm env}\rho_{\rm env} \Psi(e) \over \rho_p r_p }}\right] {N_{\rm enter}(r_p)}\\
= {\sqrt {2 \pi}}n_{\rm ISO}(r_p) \tau_{\rm env} R_{\rm env}^2 {u_{\rm esc}^2 \over \sigma} 
 \left[1-e^{-{{u_{\rm esc}^2} \over \sigma^2}{3 \over 8}{C_D R_{\rm env}\rho_{\rm env} \Psi(e) \over \rho_p r_p }}\right] {N_{\rm enter}(r_p)}, 
 \label{Nrptclargeelli}
\end{split}
\end{equation}
that for small arguments of the exponential becomes
\begin{equation}
\begin{split}
N(r_p) = {3 \over 8} {\sqrt {2 \pi}}{C_D}  
{R_{\rm env}\rho_{\rm env} \Psi(e) \over \rho_p r_p } n_{\rm ISO}(r_p)\sigma \pi R_{\rm env}^2 \tau_{\rm env} \bar\Theta_s^2 
= {9 \over 32} {\sqrt {2 \pi}}{C_D}\Psi(e) {n_{\rm ISO}(r_p)\sigma \over \rho_p r_p }M_{\rm env} \tau_{\rm env} \bar\Theta_s^2.
\label{NrpISOfinalelli}
\end{split}
\end{equation}
In the expression above, $n_{\rm ISO}(r_p)\sigma \pi R_{\rm env}^2$ is the rate at which interstellar objects cross the environment, that multiplied by $\tau_{\rm env}$ gives a total number. The other factors take into account gravitational focussing and the trapping efficiency due to gas drag. The latter provides a factor of $R_{\rm env}\rho_{\rm env}$, which allows to calculate something similar to a dust-to-gas mass ratio but with the mass of the trapped objects (see Figure \ref{isogasratio}), 
\begin{equation}
\begin{split}
{N(r_p) m_p \over M_{\rm env}}= {3 \over 8} {\sqrt {2 \pi}}{C_D}  
 \Psi(e) \pi r_p^2 n_{\rm ISO}(r_p)\sigma \tau_{\rm env} \bar\Theta_s^2.
\label{isogasratioeq}
\end{split}
\end{equation}
We can also calculate the fraction of interstellar objects that are trapped, out of the total number that are in the cloud at any given time, 
\begin{equation}
\begin{split}
{N(r_p) \over n_{\rm ISO}(r_p) {4\over3}\pi R_{\rm env}^3}= {9 \over 32} {\sqrt {2 \pi}}{C_D}  
 \Psi(e) {\sigma \tau_{\rm env} \over R_{\rm env}} {\rho_{\rm env}R_{\rm env}\over \rho_p r_p}\bar\Theta_s^2.
\label{trappednontrappedratio}
\end{split}
\end{equation}

We will now evaluate Equations \eqref{Nrptclargeelli}, \eqref{isogasratioeq} and \eqref{trappednontrappedratio} using the environmental parameters ($R_{\rm env}$, $\rho_{\rm env}$, $u_{\rm esc}$, $\tau_{\rm env}$) from Table 1 (assuming the cloud is made of H$_2$ gas) and the gas drag coefficient $C_D$ calculated in Section \ref{appendixA} and listed in Table A1. Regarding the velocity dispersion, $\sigma$, interstellar object observations are too limited to be able to constrain their velocity distribution and how it compares to stars-forming-regions. As discussed in Section \ref{ISO capture ISM}, Mackereth et al. (\citeyear{2019MNRAS.489..176M}) found that for the young stars in the Galaxy with ages $<$ 2 Gyr, their velocity dispersions are $\sigma_R \sim$ 30 km/s and $\sigma_z \sim$10 km/s. Because these young stellar systems are thought to be the main source of interstellar planetesimals, we adopt a velocity dispersion of $\sigma$ = 30 km/s. Note that this underestimates the width of the velocity distribution of the interstellar objects, ignoring their ejection velocity from their young parent stellar systems, expected to be in the 1--10 km/s range. 

The remaining factor that we need to evaluate Equations \eqref{Nrptclargeelli}, \eqref{isogasratioeq} and \eqref{trappednontrappedratio} is the number density of interstellar planetesimals, $n_{\rm ISO}(r_p)$, that we discuss below.

\subsection{Number density of interstellar planetesimals}
\label{Number density of interstellar planetesimals}
We now calculate $n_{\rm ISO}(r_p)$ for planetesimals in the size ranges of 50 cm--5 m, 5 m--50 m, 50 m--500 m, 500 m--5 km, and 5 km--50 km, that we will refer to as $N_{\rm R_1<r_p<R_2}$, with the corresponding $R_1$ and $R_2$ values. We will consider two scenarios. 

In the first case, described in Section \ref{NOumuamua}, we assume that the number density of interstellar planetesimals larger than 1I/'Oumuamua's is $2\cdot10^{15} {\rm pc}^{-3}$, based 1I/'Oumuamua's detection and following Do et al. (\citeyear{2018ApJ...855L..10D}). 

However, Do et al. (\citeyear{2018ApJ...855L..10D}) estimate requires an ejection of about 50 M$_{\oplus}$ of solids per stellar system, higher, for example, than Adams \& Spergel (\citeyear{2005AsBio...5..497A}) estimate of $\gtrsim$ M$_{\oplus}$ per star (see also Grishin et al. \citeyear{2019MNRAS.487.3324G}). Indeed, Do et al. (\citeyear{2018ApJ...855L..10D}) estimate is higher than what the protoplanetary reservoirs can provide  (Raymond et al. \citeyear{2018MNRAS.476.3031R}; Moro-Mart{\'{\i}}n \citeyear {2018ApJ...866..131M}). Other mechanisms that will contribute to the number density of interstellar objects beyond the protoplanentary disk stage include planet-planet scattering (Adams \& Spergel \citeyear{2005AsBio...5..497A}) and the release of exo-Oort cloud objects due to stellar-mass loss during the post-main sequence evolution (Moro-Mart{\'{\i}}n \citeyear{2019AJ....157...86M}). However, the contribution from these latter mechanisms ultimately depend on the initial overall circumstellar reservoir. To take into account all these potential contributors, in Section \ref{protoplanetaryejection} we adopt for the number density of interstellar objects the value that would be expected from the ejection of the entire circumstellar disk reservoir, following Moro-Mart{\'{\i}}n (\citeyear {2018ApJ...866..131M}). 
 
\subsubsection{Number density of planetesimals inferred from the detection of 1I/'Oumuamua}
\label{NOumuamua}
\begin{figure}
\begin{center}
\includegraphics[width=9cm]{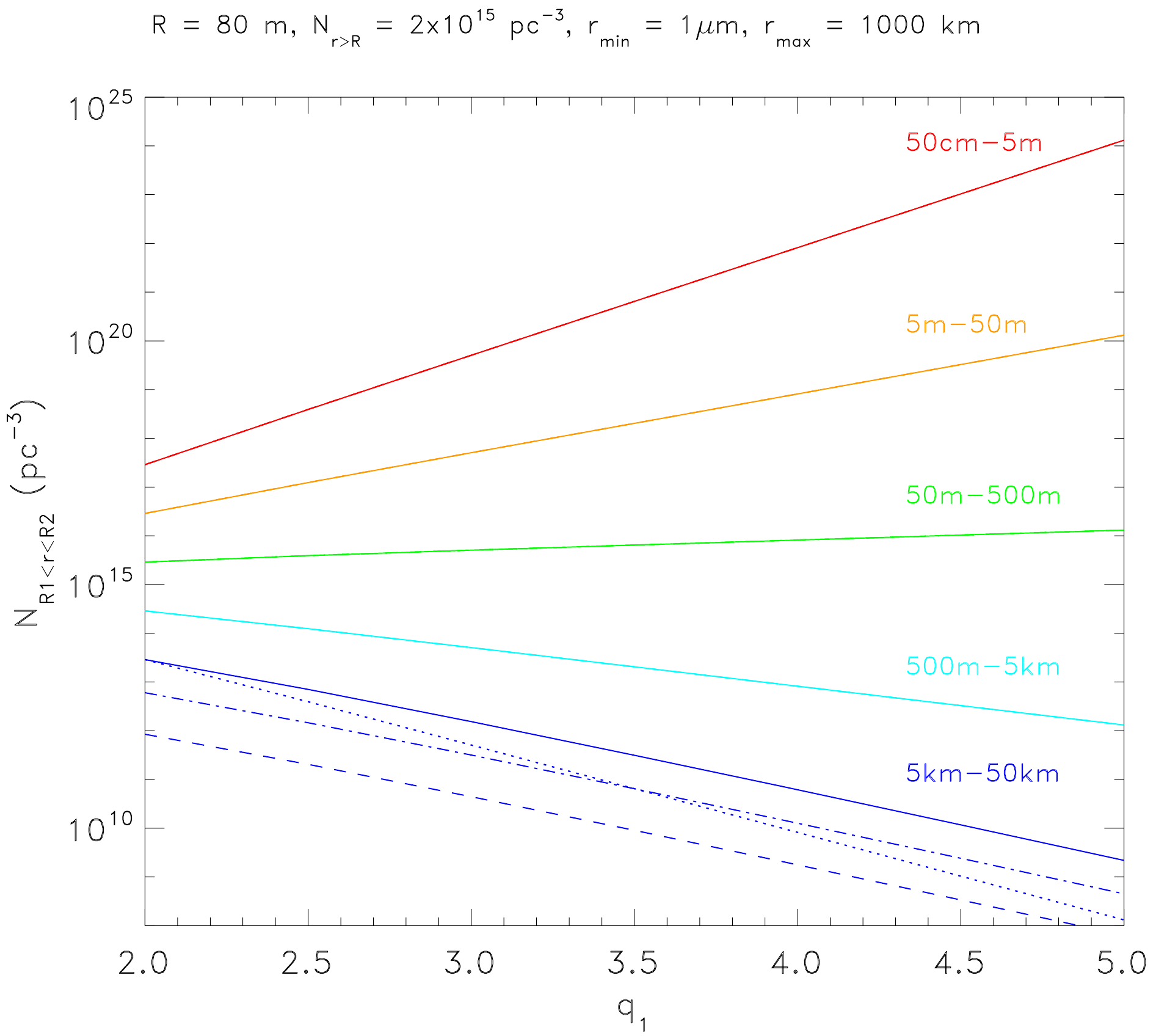}
\end{center}
\caption{Number density of interstellar objects in different size ranges, $N_{\rm R_1<r_p<R_2}$, as a function of the adopted value for the power-law index $q_1$,  assuming that the cumulative number density of interstellar planetesimals with radius similar or larger than 1I/'Oumuamua is $N_{\rm R\geqslant R_{\rm Ou}} = 2\cdot10^{15} {\rm pc}^{-3}$ (Do et al. \citeyear{2018ApJ...855L..10D}), with $R_{\rm Ou} \sim$ 80 m. The different colors correspond to the different size ranges as labeled. For the 5 km--50 km size range, the results also depend on $q_2$ and $r_b$, with the line types corresponding to the following power-law parameters:  $q_{\rm 2}$ = 2 and $r_{\rm b}$ = 3 km (solid),  $q_{\rm 2}$ = 3.5 and $r_{\rm b}$ = 3 km (dashed-dotted), $q_{\rm 2}$ = 6 and $r_{\rm b}$ = 3 km (dashed),  $q_{\rm 2}$ = 2--6 and $r_{\rm b}$ = 90 km (dotted).}
\label{N_all}
\end{figure}

Under this scenario, we adopt the cumulative number density of interstellar planetesimals inferred from 1I/'Oumuamua's detection, $N_{\rm R\geqslant R_{\rm Ou}} = 2\cdot10^{15} {\rm pc}^{-3}$ (Do et al. \citeyear{2018ApJ...855L..10D}), with $R_{\rm Ou} \sim$ 80 m. From this value we can estimate the number density of interstellar objects in the size ranges considered,  $N_{\rm R_1<r_p<R_2}$, to be used as $n_{\rm ISO}(r_p)$, by adopting the size distribution used in Moro-Mart\'{\i}n et al. (\citeyear{2009ApJ...704..733M}),
\begin{equation}
\begin{split}
n(r) \propto r^{-q_1}~{\rm if}~r_{min}<r<r_b\\
n(r) \propto r^{-q_2}~{\rm if}~r_b<r<r_{max}\\
q_1 = {\rm 2.0,~2.5,~3.0,~3.5}\\
q_2 = {\rm 3,~3.5,~4,~4.5,~5}\\
r_b = {\rm 3~km, 30~km,~90~km}\\
r_{max} \approx~{\rm1000~km~and}~r_{min} \approx~{\rm1}~\mu{\rm m}.\\
\end{split}
\label{2s-sizedist}
\end{equation}
The above distribution is based on solar system observations and on accretion and collisional models (Bottke et al. \citeyear{2005Icar..175..111B} ; Kenyon et al.  \citeyear{2008ssbn.book..293K}; Bernstein et al. \citeyear{2004AJ....128.1364B}; Fuentes, George \& Holman \citeyear{2009ApJ...696...91F}; Fraser \& Kavelaars \citeyear{2009AJ....137...72F}; Lamy et al. \citeyear{2004come.book..223L}). The results for $N_{\rm R_1<r_p<R_2}$ are shown in Figure \ref{N_all} as a function of the adopted value for $q_1$, with the different colors corresponding to the different size ranges. For the 5 km--50 km size range, the results also depend on $q_2$, and $r_b$ (as illustrated by the dotted and the dashed lines). As expected, the number density of planetesimals depend sensitively on the parameters chosen for the size distribution, lying within the ranges shown in the first row of Table 2. The values corresponding to an equilibrium size distribution are also shown in Table 2. Details on the calculations can be found in Appendix C.

\subsubsection{Number density of planetesimals in the interstellar medium expected from protoplanetary disk ejection}
\label{protoplanetaryejection}
\begin{figure}
\begin{center}
\includegraphics[width=9cm]{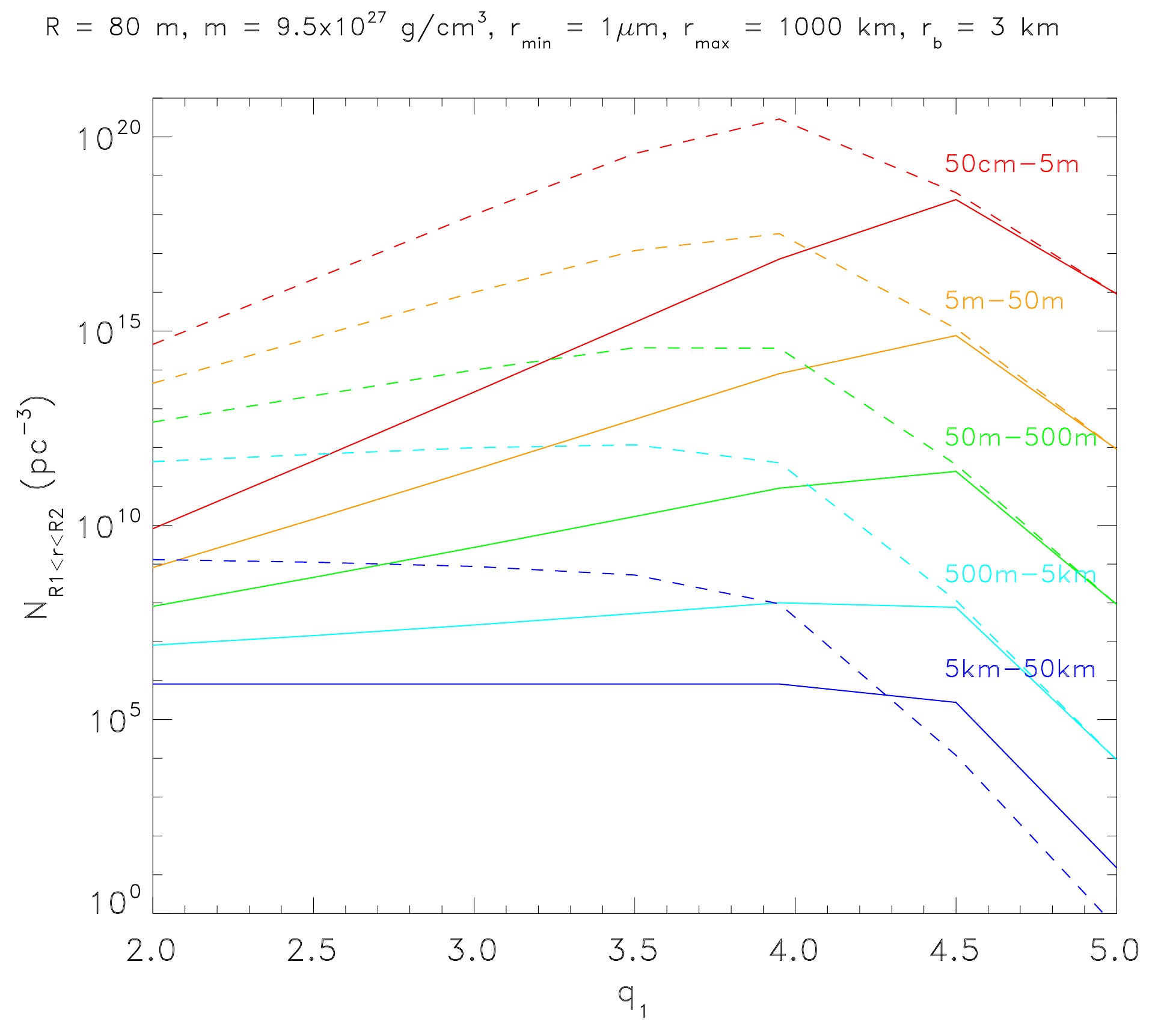}
\end{center}
\caption{Number density of interstellar objects in different size ranges, $N_{\rm R_1<r_p<R_2}$, as a function of the adopted value for the power-law index $q_1$, assuming that the total mass density of planetesimals in the interstellar medium is m$_{\rm total}$ = 9.5$\cdot$10$^{27}$ g$\cdot$pc$^{-3}$, from Moro-Mart\'{\i}n (\citeyear{2018ApJ...866..131M}; see Section \ref{protoplanetaryejection}). The different colors correspond to the different size ranges as labeled and the different line types to two different values of the power-law index $q_2$,  $q_2$=2 (solid lines) and $q_2$ = 6 (dotted lines). For this panel, we adopt r$_b$ = 3 km. Details on how these values are derived can be found in Appendixes C and D.}
\label{smallN3}
\end{figure}

\begin{figure}
\begin{center}
\includegraphics[width=9cm]{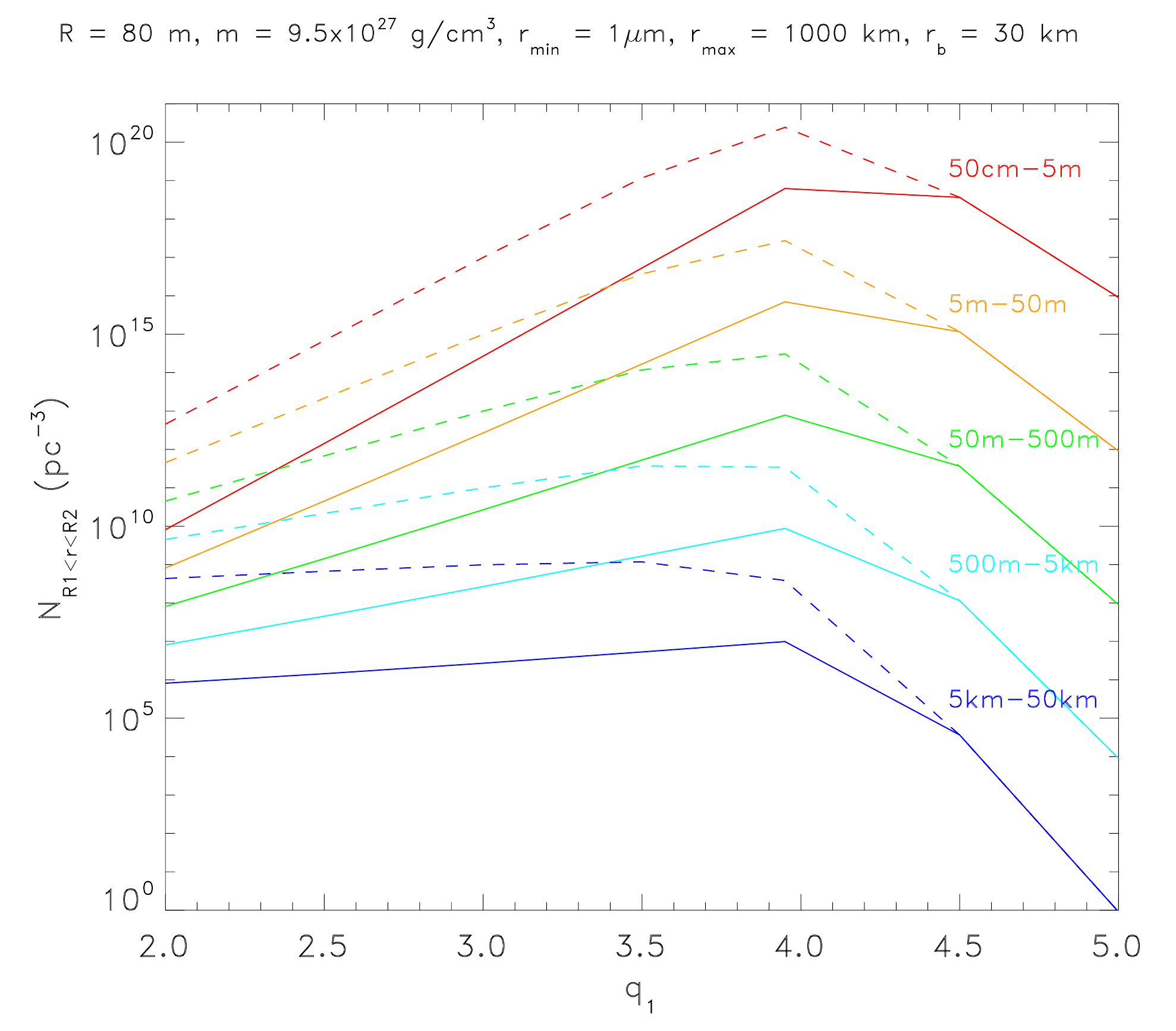}
\end{center}
\caption{Same as Figure \ref{smallN3} for r$_b$ = 30 km.}
\label{smallN30}
\end{figure}

\begin{figure}
\begin{center}
\includegraphics[width=9cm]{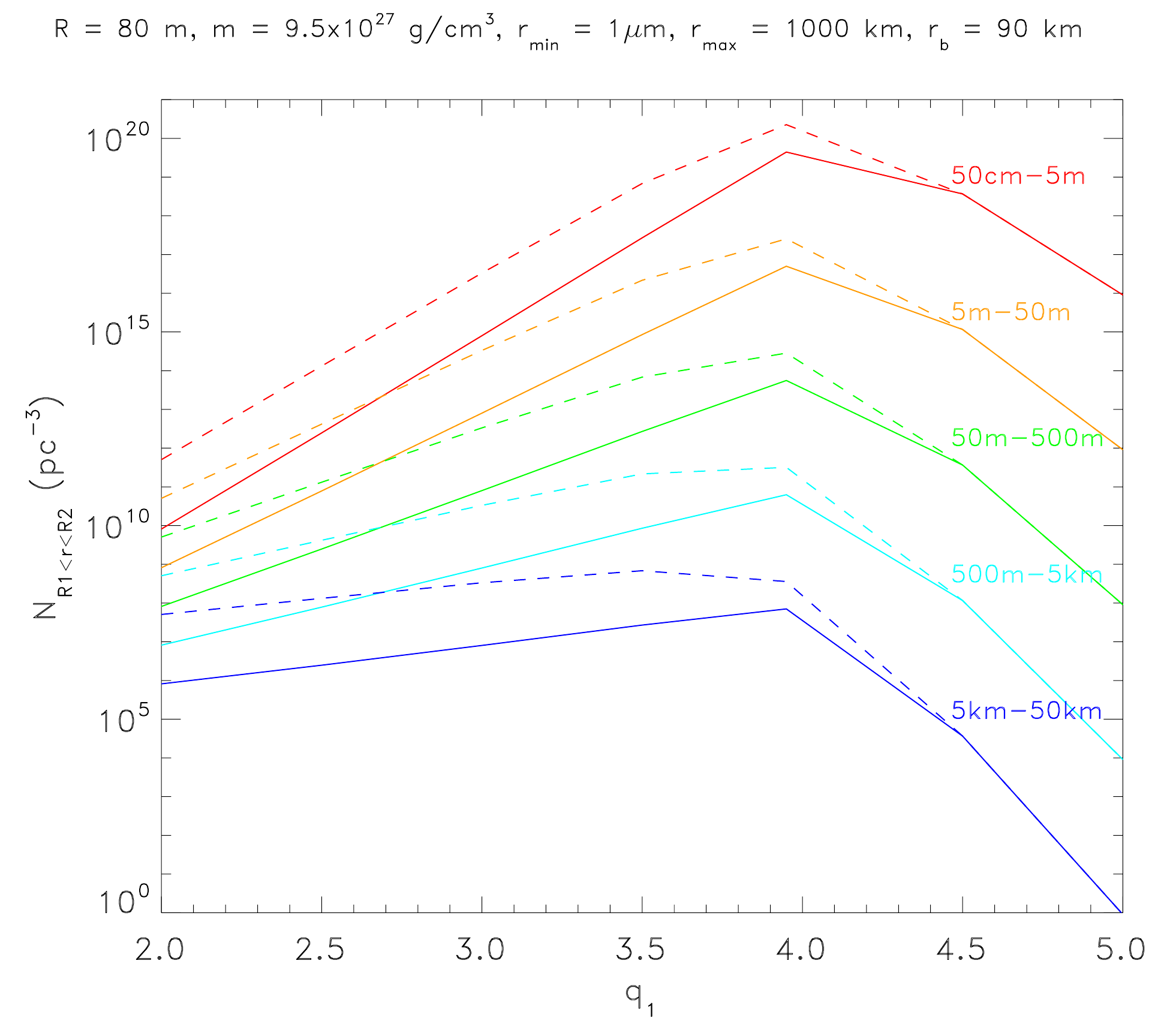}
\end{center}
\caption{Same as Figure \ref{smallN3} for r$_b$ = 90 km.}
\label{smallN90}
\end{figure}

Under this scenario, we assume that the interstellar medium is populated with planetesimals ejected from circumstellar protoplanetary disks around single stars and wide binaries, and from circumbinary protoplanetary disks around tight binaries. In terms of the expected mass density of planetesimals, Moro-Mart\'{\i}n (\citeyear{2018ApJ...866..131M}) found that 2.7$\cdot$10$^{27}$ g$\cdot$pc$^{-3}$ would be expected to originate from circumbinary disks around tight binaries\footnote{Following Jackson et al. (\citeyear{2018MNRAS.477L..85J}), this calculation assumes a binary fraction of 0.26, a disk-to-stellar system mass ratio of ${M_{disk} \over M_{sys}}$ = 0.1M$_{sys}$, that 10\% of this material migrates in, crossing the unstable radius at which point objects are ejected, and a  gas-to-dust ratio of 100:1 (for details see Section 3 of Moro-Mart\'{\i}n \citeyear{2018ApJ...866..131M}).}, and 6.7$\cdot$10$^{26}$ g$\cdot$pc$^{-3}$ or 6.8$\cdot$10$^{27}$ g$\cdot$pc$^{-3}$ from single stars and wide binaries\footnote{These two estimates of the contribution from single stars and wide binaries  assume that 74\% stars are in this category, a disk-to-stellar mass ratio of ${M_{disk} \over M_*}$ = 0.01, and a gas-to-dust ratio of 100:1. The  6.7$\cdot$10$^{26}$ g$\cdot$pc$^{-3}$ estimate corresponds to the assumption that only planet bering contribution, adopting a planet fraction of 3\% for K2--M stars and 20\% for A-K-K2 stars. The 6.7$\cdot$10$^{27}$ g$\cdot$pc$^{-3}$ estimate corresponds to the assumption that all stars contribute. For details see Section 3 of Moro-Mart\'{\i}n (\citeyear{2018ApJ...866..131M}).}. In this latter case, the overall contribution from single and binary stars would result in a total mass density of planetesimals in the interstellar medium of m$_{\rm total}$ = 9.5$\cdot$10$^{27}$ g$\cdot$pc$^{-3}$. Adopting this overall contribution and the size distribution in Equation \eqref{2s-sizedist}, we can estimate the number of interstellar objects in the size ranges considered. The details of this calculation can be found in Appendix  D. The results for $N_{\rm R_1<r_p<R_2}$ are shown in Figures \ref{smallN3}--\ref{smallN90}. The number density of planetesimals depend sensitively on the parameters chosen for the size distribution, lying within the ranges shown in the third row of Table 2. If we were to adopt a disk-to-star mass ratio of 0.1 instead of 0.01 (which could be consistent with the disk observations given the uncertainties in the dust opacities), the values in Table 2 would increase by a factor of 10, as the total mass density of interstellar objects would increase by a factor of 10, i.e. m$_{\rm total}$ = 9.5$\cdot$10$^{28}$ g$\cdot$pc$^{-3}$ (the corresponding results are shown in the fourth row of Table 2).

\section{Total number of captured planetesimals: Molecular cloud}
\label{molecularcloud}

\subsection{Previous study}
Pfalzner \& Bannister (\citeyear{2019ApJ...874L..34P}) argue that during the molecular formation process, the number density of interstellar objects inside the molecular cloud could be enhanced with respect to that in the interstellar medium ($10^{15}$ pc$^{-3}$, from Do et al. (\citeyear{2018ApJ...855L..10D})) by the same factor as the gas ($10^4$--$10^6$), and that this could lead to a number density as high as $10^{19}$ pc$^{-3}$.  They also argue that because in the molecular cloud clumps that lead to star formation the gas is further enhanced by a factor of $10^2$--$10^3$, the interstellar objects could experience a similar enhancement factor, resulting in a number density of interstellar objects that could be as high as $10^{22}$ pc$^{-3}$. 

In spite of the arguments above, for the number density of interstellar objects inside the cloud clumps, Pfalzner \& Bannister (\citeyear{2019ApJ...874L..34P}) adopt the baseline interstellar value of $10^{15}$ pc$^{-3}$ (from Do et al. \citeyear{2018ApJ...855L..10D}). This is not necessarily a conservative assumption for two reasons: first, the value of $10^{15}$ pc$^{-3}$ is uncertain because it is based on only one detection, assumes an isotropic distribution and, as we mentioned in Section \ref{Number density of interstellar planetesimals}, it is difficult to find reservoirs that can account for such a high number density of interstellar objects; second, as Pfalzner \& Bannister (\citeyear{2019ApJ...874L..34P}) point out, interstellar objects will have a velocity distribution but only those in the low-velocity range will be incorporated to the cloud (they point out that if interstellar objects have a velocity distribution resembling that of young stars $<$ 2 Gyr, assumed to be their primary source, only $\sim$ 1--4\% will have relative velocities $<$ 10 km/s).

Pfalzner \& Bannister (\citeyear{2019ApJ...874L..34P}) assume that the volume of the clumps, i.e. the volume of space in the molecular cloud that each star draws material from, is $V = {4 \over 3N} \pi R_{\rm cluster}^3$, where $N$ is the number of stars in the cluster and $R_{\rm cluster}^3$ is the radius of the cluster, with N = 100--5000 and $R_{\rm cluster}$  of 0.3--0.4 pc, and 1--1.2 pc, respectively\footnote{This assumes an average stellar mass of 0.5 M$_{\odot}$, a star-formation efficiency of 10--30\%, and a mass-radius relationship following log($M_{\rm cluster}) = 3.42\pm0.01+(1.67\pm0.025){\rm log}R_{\rm cluster}$}. They also assume that only 10--30\% of the material contained within that volume contributes to the stellar system formation, drawing an analogy with the the star-formation efficiency\footnote{For example, the {\it Core to Disks Spitzer} Legacy survey allowed to compare the mass in young stellar objects to the cloud mass and found that, over the last 2 Myr, corresponding to the lifetime of the IR excess emission this survey was sensitive to, the star-formation efficiency ranges from 3--6\%, leading to 15--30\% over the lifetime of the cloud (Evans et al. \citeyear{2009ApJS..181..321E}).}. They further assume that only 1\% the interstellar material that contributes to the stellar system formation ends up incorporated to the protoplanetary disk, escaping accretion onto the star, drawing an analogy of what happens to the gas as the protoplanetary disk forms (resulting in a disk-to-star mass ratio of 0.01). Finally, then assume that only 10\% of the interstellar objects incorporated to the disk survive further loss due to processes like erosion or volatization. 

Pfalzner \& Bannister (\citeyear{2019ApJ...874L..34P})  find that approximately 10$^{6}$ objects of 1I/'Oumuamua's size could be incorporated to each stellar system as a result of the star formation process, while in the larger size range their estimate is $\sim$ 10$^{4}$--10$^{5}$ for objects with a diameter of $\sim$1 km, and $\sim$ 10$^{3}$ for a diameter of $\sim$100 km. They note that these values could be 10$^{4}$--10$^{6}$ times higher if in the molecular cloud forming process, the interstellar objects were to increase their concentration with respect to the interstellar medium by the same degree as the gas. 

\subsection{Our estimate}
\subsubsection{Number of interstellar objects expected to be captured by the entire molecular cloud}
\label{entiremolecularcloud}
As Pfalzner \& Bannister (\citeyear{2019ApJ...874L..34P}) point out, their results rely on the assumption that the interstellar objects have small velocities relative to the molecular clouds. To overcome this caveat, in our approach we will not assume that the number density of interstellar objects that are trapped inside the molecular cloud is initially the same as in the interstellar medium. At any given time, there will be interstellar objects crossing the cloud, but not necessarily on bound orbits and with small relative velocities. Instead, using the formalism developed above, we calculate the number of interstellar objects that could be captured by gas drag inside the molecular cloud. These objects will indeed have small relative velocities. As discussed in Section \ref{totalnumber}, the number of interstellar objects that cross each environment during its lifetime is $N_{\rm enter}(r_p) = n_{\rm ISO}(r_p) \pi b_{\rm max}^2\langle u \rangle\tau_{\rm env}$. The capture probability of the interstellar objects has been calculated in Sections \ref{capturerates} and \ref{improvedcapturerate}, where the condition for capture is given by how the energy dissipated by gras drag compares to the initial kinetic energy of the objects, following Grishin et al. (\citeyear{2019MNRAS.487.3324G}).  This led to  Equation \eqref{Nrptclargeelli} that we will now use, where for $n_{\rm ISO}(r_p)$ we adopt the values calculated in Section \ref{Number density of interstellar planetesimals} and listed in Table 2 as $N_{\rm R_1<r_p<R_2}$ for each of the size ranges.  The molecular cloud properties adopted are those in Table 1. Table 3 lists the resulting number of interstellar objects that are expected to be captured by gas drag in a molecular cloud, for different assumptions of their background number density.  The resulting values show a wide range because of the uncertainty in the size distribution of the interstellar objects. The values corresponding to an equilibrium size distribution (q$_1$ = $q_2$ = 3.5) are also shown in the table. The calculation assumes a value of $C_D$ = 2. However, this is an approximation because $C_D$ depends on velocity. In this particular case, Table A1 shows that $C_D$ can be as high as 11 for the slowest particles ($\sim$ 0.1 km/s). Because the number of objects captured is approximately proportional to $C_D$ (see Equation \ref{NrpISOfinalelli}), this means that the number of particles captured in those size ranges will be higher because of the  enhanced contribution from the particles in the very low-velocity end due to their increased $C_D$, but the enhancement will be modest.

\subsubsection{Number of captured interstellar objects expected to be incorporated to each protoplanetary disk at formation}
\label{eachdiskinMC}
The values listed in the top part of Table 3 correspond to the number of objects captured in the molecular cloud as a whole. We are now interested in estimating how many might be incorporated to each protoplanetary disk at formation. As Pfalzner \& Bannister (\citeyear{2019ApJ...874L..34P}), we assume that a molecular cloud leads to the formation of a cluster of stars. The volume of space in the molecular cloud that each star draws material from is $V = {4 \over 3N} \pi R_{\rm cluster}^3$, where $N$ is the number of stars in the cluster and $R_{\rm cluster}$ is the radius of the cluster, with N = 100--5000 and $R_{\rm cluster}$  of 0.3--0.4 pc, and 1--1.2 pc, respectively. As  Pfalzner \& Bannister (\citeyear{2019ApJ...874L..34P}), we further assumed that 30\% of the material contained within that volume contributes to the stellar system formation (30\% star-formation efficiency), that 1\% of the interstellar objects incorporated to the systems end up in the disk (based on the typical disk-to-star mass ratio of 0.01, assuming the interstellar objects behave similar to the gas), and that 10\% of the objects incorporated to the disk survive  erosion or volatization. Under this scenario, we get that the number of interstellar objects that could be incorporated to each protoplanetary disk in its process of formation is a factor of ${3\cdot10^{-4} \over N}\left({R_{\rm cluster} \over R_{\rm env}}\right)^3$  smaller than the number of interstellar objects trapped in the entire cloud, listed in the top part of Table 3. These factors are 3$\cdot$10$^{-12}$ and 2$\cdot$10$^{-12}$, for N = 100 and 5000, respectively. The results for N = 100 are listed in the bottom part of Table 3 and will be discussed in Section \ref{dissmolecularcloud}. For comparison, if we assume an overall star formation efficiency of 1\% for the molecular cloud, i.e. that 1\% of the cloud mass is converted into stars, the number of interstellar objects trapped per solar mass would be the values listed for $N(r_p)$ at the top of Table 3 multiplied by a factor of ${10^{-2}\over M_{\rm env}(M_{\odot})} = 1.4\cdot10^{-7}$, where  $M_{\rm env} = 6.9E4 M_{\odot}.$
 
 \section{Total number of captured planetesimals: pre-stellar kernel and protoplanetary disk}
\label{kernelmmsn}

We now describe another two capture scenarios that are relevant during the early stage of star and planet formation:  the trapping of interstellar objects by gas drag in a pre-stellar core kernel and in a protoplanetary disk. We follow the same calculation we did above but for the corresponding environmental parameters listed in Table 1. The estimated number of interstellar objects expected to be trapped in these environments is listed in Tables 4 and 5, assuming a gas drag coefficient of $C_D$ = 2. Table A1 indicates that for the pre-stellar kernel environment and for the range of particle sizes and velocities considered, $C_D$ = 2--9, while for the protoplanetary disk environment (listed under the entry DISK-30AU in Table A1) it can be up to 23 for the slowest particles ($\sim$ 0.1 km/s) with radius 50 cm--50 m; because $N(r_p) \propto C_D$ (Equation \ref{NrpISOfinalelli}), this means that the number of particles captured in those size ranges will be higher from an enhanced contribution from the particles in the very low-velocity end, due to their increased $C_D$, but the enhancement will be modest.

In the protoplanetary disk calculation above we assumed that the swept up column is that along the scale height so instead of using $R_{\rm env}\rho_{\rm env}$ in Equation \eqref{Nrptclargeelli} we use $h\rho_{\rm env}$. Following Grishin et al. (\citeyear{2019MNRAS.487.3324G}), for the mass surface density we adopted the flared disk model in Chiang \& Goldreich (\citeyear{1997ApJ...490..368C}), $\Sigma_g(r) = \Sigma_{g,0}({r \over au})^{-{3 \over 2}}$, with $\Sigma_{g,0} = 2\cdot 10^3$ g cm$^{-2}$ (close the the value $\Sigma_{g,0} = 2.7\cdot 10^3$ g cm$^{-2}$ from Hayashi \citeyear{1981PThPS..70...35H}). The number density that was used to calculate $C_D$ (listed as  DISK-30AU in Table A1) is derived from $\rho_g(r) = {\Sigma_g(r) \over \sqrt{2\pi} h}$ (Chiang \& Goldreich \citeyear{1997ApJ...490..368C}), where $h$ is the scale height, $h = \sqrt{c^2 r^3 \over GM_*}$ and  $c$ is the sound speed (Shakura \& Sunyaev \citeyear{1973A&A....24..337S}). The temperature is given by $T_{\rm env} = {280 \over \sqrt{r}}({L \over L_{\odot}})^{1\over 4}$.  At 30AU, we have $\rho = 1.7\cdot 10^{-14}$ g cm$^{-3}$ ($n = 5 \cdot 10^{9}$ cm$^{-3}$),  $T_{\rm env} \sim$ 50 K and $h \sim$ 3 au.

\startlongtable 
\begin{deluxetable*}{lllllllll}
\tablenum{2}
\tablecaption{Expected number density of interstellar objects, $N_{\rm R_1<r_p<R_2}$, for different assumptions of $n_{ISO}$ and adopting the size distributions in Equation \eqref{2s-sizedist}.}
\tablewidth{0pc}
\tablehead{
\colhead{} & 
\colhead{} & 
\colhead{50cm--5m} & 
\colhead{5m--50m} & 
\colhead{50m--500m} & 
\colhead{500m--5km} & 
\colhead{5km--50km}
}
\startdata
\bf{N$_{\rm R\geqslant R_{Ou}}$ = 2$\cdot$10$^{15}$} pc$^{-3}$ & $q_1$=2--3.5; $q_2$=3--5& 3$\cdot$10$^{17}$--10$^{24}$ & 3$\cdot$10$^{16}$--10$^{20}$ & 3$\cdot$10$^{15}$--10$^{16}$ & 10$^{12}$--3$\cdot$10$^{14}$ & 5$\cdot$10$^{7}$--3$\cdot$10$^{13}$\\
 & $q_1$=$q_2$=3.5 & 6$\cdot$10$^{20}$ & 2$\cdot$10$^{18}$ & 6$\cdot$10$^{15}$ & 2$\cdot$10$^{13}$ & 6$\cdot$10$^{10}$\\
\bf{m$_{\rm total}$} = 9.5$\cdot$10$^{27}$ g$\cdot$pc$^{-3}$
& $q_1$=2--3.5; $q_2$=3--5 & 8$\cdot$10$^{9}$--2$\cdot$10$^{20}$ & 8$\cdot$10$^{8}$--3$\cdot$10$^{17}$ & 8$\cdot$10$^{7}$--3$\cdot$10$^{14}$ & 9$\cdot$10$^{3}$--10$^{12}$ & 0.4--10$^{9}$\\
\bf{m$_{\rm total}$} = 9.5$\cdot$10$^{28}$ g$\cdot$pc$^{-3}$
& $q_1$=2--3.5; $q_2$=3--5 & 8$\cdot$10$^{10}$--2$\cdot$10$^{21}$ & 8$\cdot$10$^{9}$--3$\cdot$10$^{18}$ & 8$\cdot$10$^{8}$--3$\cdot$10$^{15}$ & 9$\cdot$10$^{4}$--10$^{13}$ & 4--10$^{10}$
\enddata
\end{deluxetable*}

\startlongtable 
\begin{deluxetable*}{lllllllll}
\tablenum{3}
\tablecaption{Number of interstellar objects expected to be captured by gas drag in a molecular cloud, for different assumptions of $n_{ISO}$ and adopting the size distributions in Equation \eqref{2s-sizedist}\tablenotemark{a}.}
\tablewidth{0pc}
\tablehead{
\colhead{} & 
\colhead{} & 
\colhead{50cm--5m} & 
\colhead{5m--50m} & 
\colhead{50m--500m} & 
\colhead{500m--5km} & 
\colhead{5km--50km}
}
\startdata
\multicolumn{7}{c}{\bf{Trapped by the entire molecular cloud\tablenotemark{b}}}\\\\
\bf{Range\tablenotemark{c} of $C_D$} & & 2--11& 2--11& 2--11& 2--11& 2--11 \\
\\
\bf{Results for $C_D$ = 2} \\
\bf{N$_{\rm R\geqslant R_{Ou}}$ = 2$\cdot$10$^{15}$} pc$^{-3}$ & $q_1$=2--3.5; $q_2$=3--5 & 8$\cdot$10$^{16}$--3$\cdot$10$^{23}$ & 8$\cdot$10$^{14}$--3$\cdot$10$^{18}$ & 8$\cdot$10$^{12}$--3$\cdot$10$^{13}$ & 3$\cdot$10$^{8}$--8$\cdot$10$^{10}$ & 10$^{3}$--8$\cdot$10$^{8}$\\
 & $q_1$=$q_2$=3.5 & 2$\cdot$10$^{20}$ & 6$\cdot$10$^{16}$ & 2$\cdot$10$^{13}$ & 6$\cdot$10$^{9}$ & 2$\cdot$10$^{6}$\\
 \bf{m$_{\rm total}$} = 9.5$\cdot$10$^{27}$ g$\cdot$pc$^{-3}$
& $q_1$=2--3.5; $q_2$=3--5 & 2$\cdot$10$^{9}$--6$\cdot$10$^{19}$ & 2$\cdot$10$^{7}$--8$\cdot$10$^{15}$ & 2$\cdot$10$^{5}$--8$\cdot$10$^{11}$ & 3--3$\cdot$10$^{8}$ & 10$^{-5}$-3$\cdot$10$^{4}$\\
\bf{m$_{\rm total}$} = 9.5$\cdot$10$^{28}$ g$\cdot$pc$^{-3}$
& $q_1$=2--3.5; $q_2$=3--5 & 2$\cdot$10$^{10}$--6$\cdot$10$^{20}$ & 2$\cdot$10$^{8}$--8$\cdot$10$^{16}$ & 2$\cdot$10$^{6}$--8$\cdot$10$^{12}$ & 3$\cdot$10$^{1}$--3$\cdot$10$^{9}$ & 10$^{-4}$-3$\cdot$10$^{5}$\\
\\
\multicolumn{7}{c}{\bf{Incorporated from the cloud to each protoplanetary disk in the process of disk formation\tablenotemark{d}}}\\\\
\bf{N = 100}\\
\bf{N$_{\rm R\geqslant R_{Ou}}$ = 2$\cdot$10$^{15}$} pc$^{-3}$ & $q_1$=2--3.5; $q_2$=3--5 & 2$\cdot$10$^{5}$--9$\cdot$10$^{11}$ & 2$\cdot$10$^{3}$--9$\cdot$10$^{6}$ & 2$\cdot$10$^{1}$--9$\cdot$10$^{1}$ & 9$\cdot$10$^{-4}$--2$\cdot$10$^{-1}$ & 3$\cdot$10$^{-9}$--2$\cdot$10$^{-3}$\\
 & $q_1$=$q_2$=3.5 &6$\cdot$10$^{8}$ & 2$\cdot$10$^{5}$ & 6$\cdot$10$^{1}$ & 2$\cdot$10$^{-2}$ & 6$\cdot$10$^{-6}$\\
 \bf{m$_{\rm total}$} = 9.5$\cdot$10$^{27}$ g$\cdot$pc$^{-3}$
& $q_1$=2--3.5; $q_2$=3--5 & 6$\cdot$10$^{-3}$--2$\cdot$10$^{8}$ & 6$\cdot$10$^{-5}$--2$\cdot$10$^{4}$ & 6$\cdot$10$^{-7}$--2 & 9$\cdot$10$^{-12}$--9$\cdot$10$^{-4}$ & 3$\cdot$10$^{-17}$--9$\cdot$10$^{-8}$ \\
\enddata
\tablenotetext{a}{In the calculation of $N(r_p)$, we adopt as $r_p$ the smallest particle radius in the corresponding size bin.}
\tablenotetext{b}{See discussion in Section \ref{entiremolecularcloud}.}
\tablenotetext{c}{Indicates the range of $C_D$ values corresponding to the smallest particle radius in the corresponding size bin, for velocities of 0--30 km/s (from the Table A1 entry corresponding to the molecular cloud).}
\tablenotetext{d}{See discussion in Section \ref{eachdiskinMC}.}
\end{deluxetable*}

\startlongtable 
\begin{deluxetable*}{lllllllll}
\tablenum{4}
\tablecaption{Same as Table 3 for the pre-stellar core kernel.}
\tablewidth{0pc}
\tablehead{
\colhead{} & 
\colhead{} & 
\colhead{50cm--5m} & 
\colhead{5m--50m} & 
\colhead{50m--500m} & 
\colhead{500m--5km} & 
\colhead{5km--50km}
}
\startdata
\bf{Range\tablenotemark{a} of $C_D$} & & 2--9& 2--9& 2--9& 2--9& 2--9 \\
\\
\bf{Results for $C_D$ = 2}\\
\bf{N$_{\rm R\geqslant R_{Ou}}$ = 2$\cdot$10$^{15}$} pc$^{-3}$ & $q_1$=2--3.5; $q_2$=3--5 & 3$\cdot$10$^{6}$--10$^{13}$ & 3$\cdot$10$^{4}$--10$^{8}$ & 3$\cdot$10$^{2}$--10$^{3}$ & 10$^{-2}$--3& 5$\cdot$10$^{-8}$--3$\cdot$10$^{-2}$\\
 & $q_1$=$q_2$=3.5 & 6$\cdot$10$^{9}$ & 2$\cdot$10$^{6}$ & 6$\cdot$10$^{2}$ & 2$\cdot$10$^{-1}$ & 6$\cdot$10$^{-5}$\\
\bf{m$_{\rm total}$} = 9.5$\cdot$10$^{27}$ g$\cdot$pc$^{-3}$
& $q_1$=2--3.5; $q_2$=3--5 & 8$\cdot$10$^{-2}$--2$\cdot$10$^{9}$ & 8$\cdot$10$^{-4}$--3$\cdot$10$^{5}$ & 8$\cdot$10$^{-6}$--3$\cdot$10$^{1}$ & 9$\cdot$10$^{-11}$--10$^{-2}$ & 4$\cdot$10$^{-16}$--10$^{-6}$\\
\bf{m$_{\rm total}$} = 9.5$\cdot$10$^{28}$ g$\cdot$pc$^{-3}$
& $q_1$=2--3.5; $q_2$=3--5 & 8$\cdot$10$^{-1}$--2$\cdot$10$^{10}$ & 8$\cdot$10$^{-3}$--3$\cdot$10$^{6}$ & 8$\cdot$10$^{-5}$--3$\cdot$10$^{2}$ & 9$\cdot$10$^{-10}$--10$^{-1}$ & 4$\cdot$10$^{-15}$--10$^{-5}$\\
\enddata
\tablenotetext{a}{Indicates the range of $C_D$ values corresponding to the smallest particle radius in the corresponding size bin, for velocities of 0--30 km/s (from the Table A1 entry corresponding to the pre-stellar core kernel).}
\end{deluxetable*}

\startlongtable 
\begin{deluxetable*}{lllllllll}
\tablenum{5}
\tablecaption{Same as Table 3 for the protoplanetary disk at 30 AU\tablenotemark{a}.}
\tablewidth{0pc}
\tablehead{
\colhead{} & 
\colhead{} & 
\colhead{50cm--5m} & 
\colhead{5m--50m} & 
\colhead{50m--500m} & 
\colhead{500m--5km} & 
\colhead{5km--50km}
}
\startdata
\bf{Range\tablenotemark{a} of $C_D$} & & 2-23& 2--23& 2--23& 2--16& 2--6 \\
\\
\bf{Results for $C_D$ = 2}\\
\bf{N$_{\rm R\geqslant R_{Ou}}$ = 2$\cdot$10$^{15}$} pc$^{-3}$ &  $q_1$=2--3.5; $q_2$=3--5   			& 3$\cdot$10$^{8}$--10$^{15}$ & 3$\cdot$10$^{6}$--10$^{10}$ & 3$\cdot$10$^{4}$--10$^{5}$ & 1--3$\cdot$10$^{2}$ & 5$\cdot$10$^{-6}$--3\\
& $q_1$=$q_2$=3.5 					& 6$\cdot$10$^{11}$ & 2$\cdot$10$^{8}$ & 6$\cdot$10$^{4}$ & 2$\cdot$10$^{1}$ & 6$\cdot$10$^{-3}$\\
\bf{m$_{\rm total}$} = 9.5$\cdot$10$^{27}$ g$\cdot$pc$^{-3}$ &$q_1$=2--3.5; $q_2$=3--5
& 8--2$\cdot$10$^{11}$ & 8$\cdot$10$^{-2}$--3$\cdot$10$^{7}$ & 8$\cdot$10$^{-4}$--3$\cdot$10$^{3}$ & 9$\cdot$10$^{-9}$--1 & 4$\cdot$10$^{-14}$--10$^{-4}$\\
\bf{m$_{\rm total}$} = 9.5$\cdot$10$^{28}$ g$\cdot$pc$^{-3}$ &$q_1$=2--3.5; $q_2$=3--5
& 8$\cdot$10$^{1}$--2$\cdot$10$^{12}$ & 8$\cdot$10$^{-1}$--3$\cdot$10$^{8}$ & 8$\cdot$10$^{-3}$--3$\cdot$10$^{4}$ & 9$\cdot$10$^{-8}$--1$\cdot$10$^{1}$ & 4$\cdot$10$^{-13}$--10$^{-3}$\\
\enddata
\tablenotetext{a}{As discussed in Section \ref{kernelmmsn}, in this order-of-magnitude estimate, the properties of the environment are those characteristic of the protoplanetary disk at 30AU, with $n = 6 \cdot 10^{13}$ cm$^{-3}$ and $h = 3 \cdot 10^{14}$ cm. We assume that the swept up column is that along the scale height of the disk, $h$, as discussed in Section \ref{kernelmmsn}.}
\tablenotetext{b}{Indicates the range of $C_D$ values corresponding to the smallest particle radius in the corresponding size bin, for velocities of 0--30 km/s (from the Table A1 entry corresponding to DISK-30AU).}
\end{deluxetable*}

\startlongtable 
\begin{deluxetable*}{lllllllll}
\tablenum{6}
\tablecaption{Same as Table 5 for the protoplanetary disk at 30 AU\tablenotemark{a} in an open cluster environment.}
\tablewidth{0pc}
\tablehead{
\colhead{} & 
\colhead{} & 
\colhead{50cm--5m} & 
\colhead{5m--50m} & 
\colhead{50m--500m} & 
\colhead{500m--5km} & 
\colhead{5km--50km}
}
\startdata
\bf{Range\tablenotemark{a} of $C_D$} & & 2-23& 2--23& 2--23& 2--16& 2--6 \\
\\
\bf{Results for $C_D$ = 2}\\
\bf{N$_{\rm R\geqslant R_{Ou}}$ = 2$\cdot$10$^{15}$} pc$^{-3}$ &  $q_1$=2--3.5; $q_2$=3--5  & 2$\cdot$10$^{13}$--6$\cdot$10$^{19}$ & 2$\cdot$10$^{11}$--6$\cdot$10$^{14}$ & 2$\cdot$10$^{9}$--6$\cdot$10$^{9}$ & 6$\cdot$10$^{4}$--2$\cdot$10$^{7}$ & 3$\cdot$10$^{-1}$--2$\cdot$10$^{5}$\\
& $q_1$=$q_2$=3.5 	& 4$\cdot$10$^{16}$ & 10$^{13}$ & 4$\cdot$10$^{9}$ & 10$^{6}$ & 4$\cdot$10$^{2}$\\
\bf{m$_{\rm total}$} = 9.5$\cdot$10$^{27}$ g$\cdot$pc$^{-3}$ &$q_1$=2--3.5; $q_2$=3--5 
& 5$\cdot$10$^{5}$--10$^{16}$ & 5$\cdot$10$^{3}$--2$\cdot$10$^{12}$ & 5$\cdot$10$^{1}$--2$\cdot$10$^{8}$ & 6$\cdot$10$^{-4}$--6$\cdot$10$^{4}$ & 3$\cdot$10$^{-9}$--6\\
\bf{m$_{\rm total}$} = 9.5$\cdot$10$^{28}$ g$\cdot$pc$^{-3}$ & $q_1$=2--3.5; $q_2$=3--5
& 5$\cdot$10$^{6}$--10$^{17}$ & 5$\cdot$10$^{4}$--2$\cdot$10$^{13}$ & 5$\cdot$10$^{2}$--2$\cdot$10$^{9}$ & 6$\cdot$10$^{-3}$--6$\cdot$10$^{5}$ & 3$\cdot$10$^{-8}$--6$\cdot$10$^{1}$\\
\enddata
\tablenotetext{a}{As discussed in Section \ref{kernelmmsn}, in this order-of-magnitude estimate, the properties of the environment are those characteristic of the protoplanetary disk at 30AU, with $n = 6 \cdot 10^{13}$ cm$^{-3}$ and $h = 3 \cdot 10^{14}$ cm. We assume that the swept up column is that along the scale height of the disk, $h$, as discussed in Section \ref{kernelmmsn}.}
\tablenotetext{b}{Indicates the range of $C_D$ values corresponding to the smallest particle radius in the corresponding size bin, for velocities of 0--30 km/s (from the Table A1 entry corresponding to DISK-30AU).}
\end{deluxetable*}

\startlongtable 
\begin{deluxetable*}{llllllllll}
\tablenum{7}
\tablecaption{Results from Grishin et al. (\citeyear{2019MNRAS.487.3324G})\tablenotemark{a}}
\tablewidth{0pc}
\tablehead{
\colhead{} & 
\colhead{50cm} & 
\colhead{5m} & 
\colhead{50m} & 
\colhead{500m} & 
\colhead{5km} &
\colhead{50km}
}
\startdata
Field & 3$\cdot$10$^{12}$ & 2$\cdot$10$^{9}$ & 8$\cdot$10$^{5}$ & 8$\cdot$10$^{2}$ & 1 & 10$^{-3}$\\
Cluster & 1$\cdot$10$^{14}$ & 2$\cdot$10$^{11}$ & 9$\cdot$10$^{7}$ & 10$^{5}$ & 10$^{2}$ & 10$^{-1}$\\
\enddata
\tablenotetext{a}{Derived from Grishin et al. (\citeyear{2019MNRAS.487.3324G}) Figure 3 left, corresponding to an equilibrium size distribution ($p$=11/6) and a number density of interstellar objects of $2\cdot10^{15}~{\rm pc}^{-3}$ (shown in their figure as transparent blue and green, for a field and cluster environments, respectively). These values are to be compared to those in Tables 7 and 8 corresponding to $N_{\rm R\geqslant R_{\rm Ou}}$ =  2$\cdot$10$^{15}$ pc$^{-3}$ and an equilibrium size distribution, $q_1$ = $q_2$ = 3.5.}
\end{deluxetable*}

\section{The case of an open cluster environment}
\label{ISO capture cluster}
\subsection{Increased capture probability in an open cluster environment}
The capture efficiency of interstellar objects by a stellar system increases when considering stars in an open cluster, a common environment for stellar birth (Adams \& S9cmpergel \citeyear{2005AsBio...5..497A}). This is because the relative velocities of the stars in the cluster are smaller than the relative velocities of the stars in the field. 
Belbruno et al. et al. (\citeyear{2012AsBio..12..754B}) found that, when considering chaotic, nearly-parabolic orbits, the transfer probability of solids between the stars in an open cluster is many orders of magnitude higher than previously thought, allowing for the exchange of solid material in significant quantities between the young solar system and nearby planetary systems while the Sun was still embedded in the birth cluster. Belbruno et al. (\citeyear{2012AsBio..12..754B}) focused on the implications this has on the lithopanspermia hypothesis. Here, we address whether this exchange of interstellar planetesimals, ejected from a stellar system where planetesimal formation has already taken place, could provide potential seeds for planet formation in a nearby stellar system in the cluster. 

The characteristic escape velocities for the chaotic, nearly parabolic trajectories considered in Belbruno et al. (\citeyear{2012AsBio..12..754B}) are $\sim$0.1 km/s, resulting in incoming planetesimal velocities  dominated by the relative velocities of the stars in the cluster, assumed to be $u \sim$ 1 km/s. However, under the  scenario that we are considering in this paper, the typical ejection velocities of the objects, that will subsequently become interstellar, are significantly higher, $u \sim 6.2\pm2.7$ km/s (Adams \& Spergel \citeyear{2005AsBio...5..497A}). In the calculation of the expected number of captured objects using Equation \eqref{Nrptclargeelli}, instead of the previously assumed velocity dispersion of $\sigma$ = 30 km/s, we will use $\sigma$ = 6 km/s.

\newpage
\subsection{Number density of interstellar objects in an open cluster}
We now estimate the expected number density of interstellar objects in an open cluster, $n_{\rm ISO}(r_p)^{OC}$, to be used as $n_{\rm ISO}(r_p)$ in Equations \eqref{Nrptclargeelli} and \eqref{NrpISOfinalelli}, by taking into account the number density of stars in the cluster, their initial mass function, and the number of planetesimals expected to be ejected from each star (a function of the stellar mass).  We assume that the cluster consists on N = 100--5000 members, as clusters with this range of sizes are the birthplaces of a large fraction of stars in the Galaxy (Lada \& Lada \citeyear{2003ARA&A..41...57L}). For the cluster radius we adopt R$_{\rm cluster}$ = 1pc$\left({N \over 300}\right)^{1/2}$ (Adams \citeyear{2010ARA&A..48...47A}), leading to an average number density of stars of $n = {3N \over 4\pi R_{cluster}^3}$, resulting in 122 pc$^{-3}$ and 18 pc$^{-3}$,  for a cluster with N=100 and 5000 members, respectively. 

The overall contribution from all the stars in the open cluster to the number density of interstellar planetesimals can be estimated following the same calculation as in Section \ref{protoplanetaryejection} for $N_{\rm R_1<R<R_2}$, shown in Figures \ref{smallN3}--\ref{smallN90} and listed in Table 2, but scaling it to take into account that the initial mass function to be used needs to agree with the estimated average number density of stars in the cluster, as calculated above.  The estimate in Section \ref{protoplanetaryejection} adopted the following initial mass function from from Kroupa et al. (\citeyear{1993MNRAS.262..545K}), 
\begin{equation}
\begin{split}
n(M_*) = \xi(M_*)dM_*~~~~{\rm with,}~~~~~~~~~~~~\\ 
\xi(M_*) = 0.035 M_*^{-1.3}~~~~{\rm if~0.08} \le M_* < 0.5\\
\xi(M_*) = 0.019 M_*^{-2.2}~~~~~{\rm if~0.5} \le M_* < 1.0\\
\xi(M_*) = 0.019 M_*^{-2.7}~~~~{\rm if~1.0} \le M_* < 100,   
\end{split}
\label{eq_stellar_density}
\end{equation}
where $\xi$ is the number density of  stars per pc$^3$ and the stellar mass is in units of M$_{\odot}$.  This leads to scaling factors of $f^{\rm OC} = 8.9\cdot 10^{2}$ for N = 100 and $f^{\rm OC} = 1.3\cdot 10^{2}$ for N = 5000. We then have that the number density of interstellar objects in the open cluster is $n_{\rm ISO}(r_p)^{OC} = n_{\rm ISO}(r_p) \cdot f^{\rm OC}$.   

\subsection{Number of captured interstellar planetesimals per protoplanetary disk in an open cluster environment}

The equivalent of Equation \eqref{NrpISOfinalelli} in the cluster environment, estimating the expected number of captured interstellar planetesimals per protoplanetary disk, is given by
\begin{equation}
\begin{split}
N(r_p) = {3 \pi \over 8} {\sqrt {2 \pi}}{C_D} n_{\rm ISO}(r_p) f^{\rm OC} \tau_{\rm env} R_{\rm env}^2 u_{\rm esc} 
\left({u_{\rm esc} \over \sigma}\right)^3{R_{\rm env}\rho_{\rm env} \Psi(e) \over \rho_p r_p }.
\label{NrpISOfinalellicluter}
\end{split}
\end{equation}

The results are listed in Table 6, and are scaled from those in Table 5 by a factor of $f^{\rm OC}\left({\sigma_{field} \over \sigma_{OC}}\right)^3 = 500({30 \over 6})^3$ = 6.25$\cdot$10$^{4}$, where we have assumed $f^{\rm OC} = 500$. 

\section{Discussion}
\label{discussion}

\subsection{Molecular cloud environment}
\label{dissmolecularcloud}
 
In Section \ref{molecularcloud}, we estimated how many interstellar objects can be trapped by the molecular cloud. Assuming these objects are uniformly distributed, we followed Pfalzner \& Bannister (\citeyear{2019ApJ...874L..34P}) to get a rough estimate of how many of these trapped interstellar objects could be incorporated to each star-forming disk. We looked at the case of a cloud clump forming a star cluster with N = 100 members. If we adopt a cumulative number density of interstellar objects in the interstellar medium of $N_{\rm R\geqslant R_{\rm Ou}}$ =  2$\cdot$10$^{15}$ pc$^{-3}$ and the size distributions given by Equation \eqref{2s-sizedist}, we find that the number of trapped objects for the different particle sizes are:  
2$\cdot$10$^{5}$--9$\cdot$10$^{11}$ (50 cm--5 m),
2$\cdot$10$^{3}$--9$\cdot$10$^{6}$ (5 m--50 m), 
2$\cdot$10$^{1}$--9$\cdot$10$^{1}$ (50 m--500 m),
9$\cdot$10$^{-4}$--2$\cdot$10$^{-1}$ (500 m--5 km), and 
3$\cdot$10$^{-9}$--2$\cdot$10$^{-3}$ (5 km--50 km).
 
For comparison, Pfalzner \& Bannister (\citeyear{2019ApJ...874L..34P}), using the same range of size distributions, estimate that the number of interstellar objects incorporated to each disk at formation will be approximately 10$^{6}$ of 1I/'Oumuamua's size, $\sim$ 10$^{4}$--10$^{5}$ with a diameter of $\sim$1 km, and $\sim$ 10$^{3}$ for a diameter of $\sim$100 km. Their results are several orders of magnitude higher than our estimates. Our approach differs from Pfalzner \& Bannister (\citeyear{2019ApJ...874L..34P}) in that the number density of interstellar objects that we use in our calculation, for the different size ranges, corresponds to the objects that have been trapped from the interstellar medium by the molecular cloud, while Pfalzner \& Bannister (\citeyear{2019ApJ...874L..34P}) assumes that the number density of interstellar objects inside the cloud is the same as  in the interstellar medium. Their approach relies on the assumption that the interstellar objects have small velocities relative to the molecular clouds. We assume a  velocity dispersion of 30 km/s and in this case the relative velocities can be significant. At any given time, there will  be interstellar objects crossing the cloud, but most of them will not be trapped (see below) and will not necessarily have small relative velocities. However, we can assume that the small number of objects that are trapped will indeed have small relative velocities. We can calculate the fraction of 
interstellar objects that are trapped, out of the total number that are in the cloud at any given time, using Equation \eqref{trappednontrappedratio}: adopting a lifetime, $\tau_{\rm env}$ = 30 Myr, we get that 
${N(r_p) \over n_{\rm ISO}(r_p) {4\over3}\pi R_{\rm env}^3}= {9 \over 32} {\sqrt {2 \pi}}{C_D}  \Psi(e) {\sigma \tau_{\rm env} \over R_{\rm env}} {\rho_{\rm env}R_{\rm env}\over \rho_p r_p}\bar\Theta_s^2 = {3\cdot10^{-3}\over r_p}$, where $r_p$ is in cm.

However, Dobbs \& Pringle (\citeyear{2013MNRAS.432..653D}) note that the molecular cloud lifetime may be effectively much longer than the  $\tau_{\rm env}$ = 30 Myr value adopted above. This is because giant molecular clouds may dissolve by breaking up into small fragments, but these fragments will maintain their molecular cloud identity during at least a rotation period or even many rotation periods around the Galaxy, reforming giant molecular clouds in the spiral arms that will then fragment again after $\tau_{\rm env}\sim $ 30 Myr. In this case, the number of interstellar objects that are trapped by the cloud fragment can be significantly increased because of its increased lifetime.

As Pfalzner \& Bannister (\citeyear{2019ApJ...874L..34P}) point out, this order-of-magnitude calculation relies heavily on the assumption that the interstellar objects behave similarly to the gas in the cloud- and star-formation processes, but this is only valid if these processes are dominated by gravitational forces and not by gas dynamics, because the interstellar objects and the gas are largely decoupled. In an interesting new study using numerical simulations, Pfalzner et al. (\citeyear{2021arXiv210608580P}) confirm that  the interstellar objects follow the collapse of the gas in the molecular cloud. They also find that the number of interstellar objects bound to each future star-forming system is disproportionately  higher for the more massive systems that will eventually lead to subclusters of stars. In their low-resolution numerical simulations, each interstellar test particle represents about 10$^{10}$ interstellar objects, and this does not allow to follow the fate of smaller ensembles of interstellar objects. Due to the resolution of the simulations, it is also not possible to model how many interstellar objects will be bound to solar-type stars. But their very interesting initial proof-of-concept results indicate that the number of interstellar objects that might be incorporated to each star-forming disk can be very diverse, contrary to our naive scenario in which the interstellar objects captured by the molecular cloud are distributed uniformly among the stars in the cluster. They find that some star forming systems are very rich in bound interstellar objects while others are depleted of them. Pfalzner et al. (\citeyear{2021arXiv210608580P}) simulations assume that the initial distribution of interstellar objects are not moving relative to the cloud but, as pointed out earlier, the velocity dispersion can be significant. Future numerical simulations accounting for the velocity distribution of the interstellar objects should also consider the population of objects already trapped by the cloud, as calculated in our study.

\subsection{Protoplanetary disk environment}
\label{dissprotoplanetarydisk}
Table 5 shows the number of interstellar objects expected to be captured by gas drag in a protoplanetary disk, for different assumptions of the number density of interstellar objects and adopting the size distributions in Equation \eqref{2s-sizedist}. When adopting a cumulative number density of $N_{\rm R\geqslant R_{\rm Ou}}$ =  2$\cdot$10$^{15}$ pc$^{-3}$ and an equilibrium size distribution ($q_1$ = $q_2$ = 3.5), we find that the number of captured objects are 
6$\cdot$10$^{11}$ (50 cm--5 m), 
2$\cdot$10$^{8}$ (5 m--50 m), 
6$\cdot$10$^{4}$ (50 m--500 m), 
2$\cdot$10$^{1}$ (500 m--5 km), and 
6$\cdot$10$^{-3}$ (5 km--50 km). These results are lower than those in Grishin et al. (\citeyear{2019MNRAS.487.3324G}) for $p$ = 11/6 (corresponding also to an equilibrium size distribution), listed in Table 7 for the field environment. Our results show that a significant number of interstellar planetesimals can be incorporated to the protoplanetary disks by gas drag. But they also indicate that the uncertainty in the expected number of captured objects is very significant. This is due to the wide range of possible size (mass) distributions for interstellar objects in Equation \eqref{2s-sizedist} (this uncertainty is reflected in the wide range of values listed in Table 5 under $q_1$=2--3.5; $q_2$=3--5),  and to the uncertainties in the background number density of interstellar objects (from a high value of $N_{\rm R\geqslant R_{\rm Ou}}$ =  2$\cdot$10$^{15}$ pc$^{-3}$, corresponding to the number density inferred from 1I/'Oumuamua's detection, to a lower estimate labeled as m$_{\rm total}$ = 9.5$\cdot$10$^{28}$ g$\cdot$pc$^{-3}$, corresponding to what would be expected from the ejection of planetesimals from protoplanetary disks).  In the future, we will be able to narrow down these estimates as the population of interstellar objects becomes better characterized, both observationally and via simulations, in particular its background number density, size, and velocity distributions. The simulations of the latter should take into account the  ejection process of the planetesimals from their parent system and how their long journey in the Galaxy affects their initial velocity, as the interstellar planetesimal population is subject to size-dependent dynamical heating.

\subsection{Trapped interstellar objects-to-gas mass ratio}
\begin{figure}
\begin{center}
\includegraphics[width=12cm]{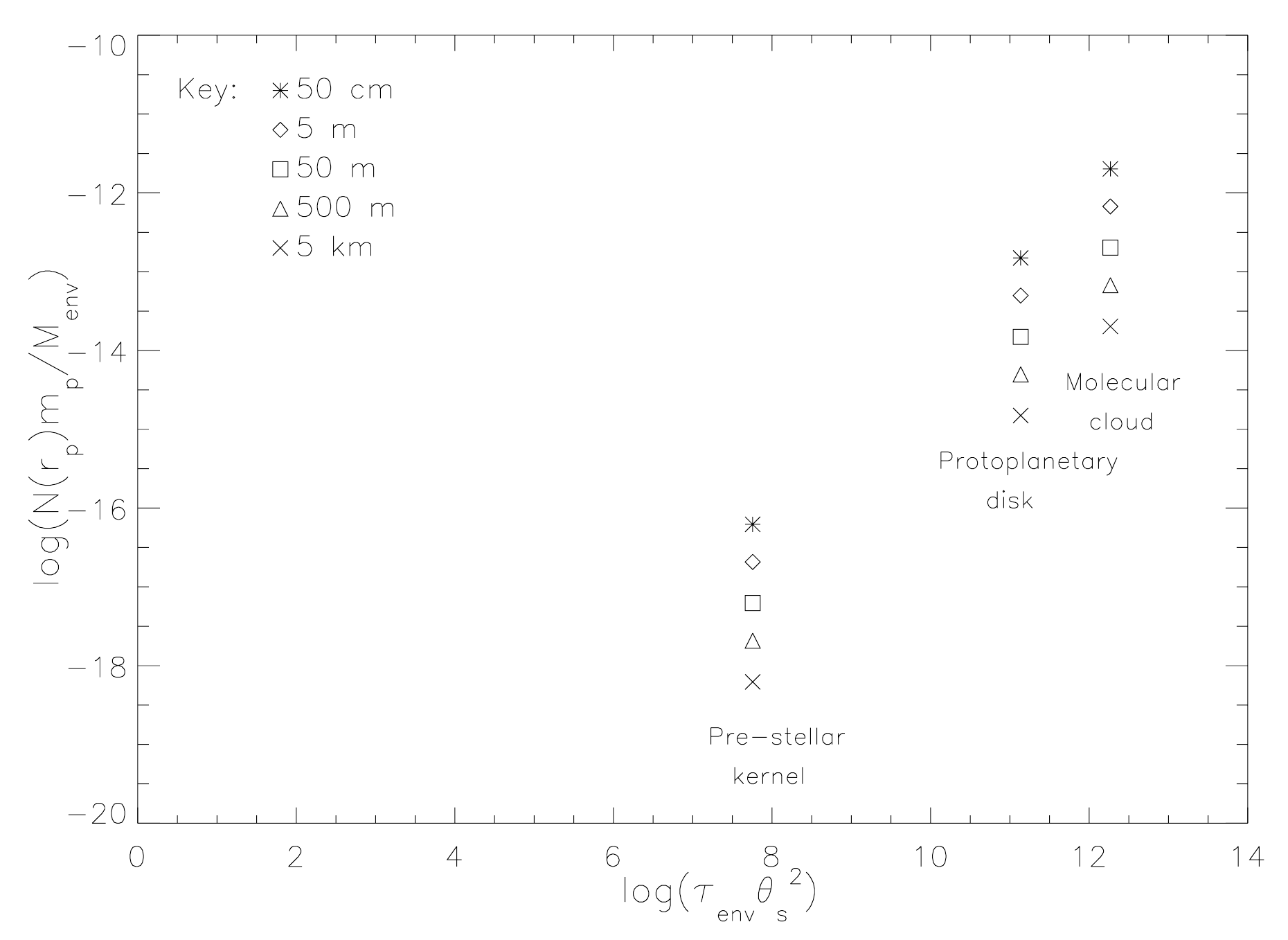}
\end{center}
\caption{Trapped interstellar objects-to-gas mass ratio for the molecular cloud, pre-stellar kernel and protoplanetary disk environments, calculated using Equation \eqref{isogasratioeq} and assuming an equilibrium size distribution for the interstellar objects.}
\label{isogasratio}
\end{figure}

Using Equation \eqref{isogasratioeq} we can calculate the mass ratio of the trapped interstellar objects to the gas in the environment, ${N(r_p) m_p \over M_{\rm env}}$. Figure \ref{isogasratio} shows the results when we adopt an equilibrium size distribution for the interstellar objects. This is just a zeroth-order estimate that assumes the interstellar objects are uniformly distributed in the environment. However, their mixing within the environment will be complex because of the dynamics (as discussed by Pfalzner et al. \citeyear{2021arXiv210608580P} in the case of a star-forming molecular cloud), because the gas does not have a uniform density, and because the size and phase space distributions of the interstellar objects are uncertain, affecting their stokes numbers and trapping efficiencies.   

\subsection{Implications}
\label{implications}
The interest of these trapped interstellar objects is that they could have sizes large enough to overcome the drift barrier, as the maximum drift speed in the protoplanetary disk is found for bodies around a meter in size, decreasing significantly for larger sizes: for planetesimals with bulk densities of 3 g$\cdot$cm$^{-3}$ located at a radial distance of 5 AU in a solar nebula, Weidenschilling (\citeyear{1977MNRAS.180...57W}) found drift timescales of O(10$^{2})$ yr, O(10$^{4})$ yr, O(10$^{5})$ yr, and O(10$^{7})$ yr, for particles with sizes of 1 m, 10 m, 100 m, and 1 km, respectively. The settling timescale into the disk mid-plane also decreases with size (Weidenschilling \citeyear{1980Icar...44..172W}; see e.g. Figure 2 in Bate \& Lor{\'e}n-Aguilar \citeyear{2017MNRAS.465.1089B}). Both processes can favor the rapid growth of these trapped planetesimals into larger bodies via the direct accretion of the sub-cm sized dust grains in the protoplanetary disk, with their more rapid settling time allowing them to get a boost in accretion growth as they settle into the mid-plane. This can lead to the growth of planetesimals by a factor of 10$^6$ in mass before the onset of collisions with other planetesimals in the disk and erosion (that start to become important at planetesimal sizes of $\sim$100 km, Xie et al. \citeyear{2010ApJ...724.1153X}). Mutual collisions between the trapped interstellar planetesimals can be neglected\footnote{This is because the planetesimals' growth rate, $t_{\rm growth} = {\bar{m} \over \dot{m}} = {\bar{m} \over \rho_{\rm dust}^{\rm disk} \pi r^2 \sigma_{\rm dust}^{\rm vel}}$, times the collisional rate of the planetesimals, $R_{\rm col} =  n \pi r^2 \sigma^{\rm vel}$, indicates that for mutual collisions to become important, i.e. $R_{\rm coll}t_{\rm growth} = {n\bar{m} \over \rho_{\rm dust}^{\rm disk}}{\sigma^{\rm vel} \over  \sigma_{\rm dust}^{vel}} = 
{N\bar{m} \over m_{\rm dust}^{\rm disk}}{\sigma^{\rm vel} \over  \sigma_{\rm dust}^{vel}} >$ 1, the number of planetesimals trapped in the disk needs to be 
$N > {m_{\rm disk}^{\rm dust} \over \bar{m}}{\sigma_{\rm dust}^{\rm vel} \over  \sigma^{vel}}$. Here, $\bar{m}$, $r$,  and $\sigma^{\rm vel}$ are the average mass, radius, and velocity dispersion of the trapped planetesimals, and $n$ and $N$ are their number density in the disk and total number, respectively; while $\rho_{\rm dust}^{\rm disk}$, $m_{\rm dust}^{\rm disk}$, and $\sigma_{\rm dust}^{\rm vel}$ are the mass density, total mass, and velocity dispersion of dust in the disk, respectively. For a 1-m sized planetesimal, assuming a dust disk with a dust mass of 10$^{-4}$M$_{\odot}$ and that ${\sigma_{\rm dust}^{\rm vel} \over  \sigma^{vel}} \sim 0.1$, collisions would become important for $N > 10^{22}$, orders of magnitude larger than the estimates in Table 2.}.

We therefore conclude that the trapping of interstellar objects could potentially provide seeds for planet formation that help overcome the meter size barrier and that this possibility should be considered in future planet formation models.  After the disk is seeded, the evolution of the objects is subject to rapid growth and collisional disruption. The actual origin of the planets and the small-body population will be a complex combination of these processes. In the solar system, for example, we have to account for the formation of a large small-body population of asteroids, comets and Kuiper belt objects; the Kuiper belt alone is thought to host of the order of 10$^5$ objects $>$ 100 km, and long-period comet observations indicate that of the order of 10$^{12}$ objects $>$ few km populate the Oort cloud. Planet-formation models are necessary to assess how many interstellar seeds should be sufficient to account for the large dynamical range of masses.

As Grishin et al. (\citeyear{2019MNRAS.487.3324G}) and Pfalzner \& Bannister (\citeyear{2019ApJ...874L..34P}) argue, this trapping mechanism could not provide the first generation of seeds but, because there is the possibility that it will have an increasingly important role later on, as the number density of interstellar planetesimals in the Galaxy increases, it might alleviate the problem of requiring all the planet-forming disks to have fine-tuned conditions to overcome the meter-sized barrier. In the case of the open cluster, for example, if we assume that only 1\% of the stars in the open cluster have undergone planet formation, leading to planetesimals ejection, we get that the expected number of  ejected objects that can end up captured by a neighboring protoplanetary system is 1\% of those listed in Table 6. The expected number of trapped objects can remain very significant, i.e. a small percentage of stars in the cluster that have successfully undergone planet formation can provide a large number of interstellar seeds. 

In this work, we have adopted the number density of interstellar objects based on observations of 1I/'Oumuamua and based on the estimated mass density expected from protoplanetary disk ejection. However, we still lack a detailed understanding of sources and sinks, including dynamical heating, that lead to the observed number density distributions (in phase space and size) of these interstellar population; this will be the subject of future work.

Another aspect that needs to be taken into account is that the porosity of the objects we are considering here can be very significant. In the solar system, the most primitive dust "aggregates" have  porosities of 60\%--70\%. More fluffy, fractal-like objects, like those expected to formed in protoplanetary disks (Okuzumi et al. \citeyear{2012ApJ...752..106O}; Kataoka et al. \citeyear{2013A&A...557L...4K}) would be trapped more easily because of their low mass per unit area and therefore our total capture rates for porous objects may increase by an order of magnitude or more. In the context of the suggestion that 1I/'Oumuamua might be a very low-density object (Moro-Mart{\'\i}n \citeyear{2019ApJ...872L..32M}), in future work we will consider this  issue of the effect of the porosity in the capture of interstellar objects. 

We  also need to take into account how the evaporation, processing, accretion and sputtering of interstellar objects in the different phases of the interstellar medium can affect  the population of planetesimals. This is of interest regardless of the nature of the objects, but is particularly important if they are highly porous or are a fragment of H$_2$ ice (Seligman \& Laughlin \citeyear{2020ApJ...896L...8S}) or N$_2$ ice (Jackson \& Desch \citeyear{2021JGRE..12606706J}), as it has been proposed for 1I/'Oumuamua.  And generally, for all interstellar objects originating from planetary systems, it is important to understand how their ejection and entry (with a wide range of dynamical histories) can alter their physical properties (Raymond et al. \citeyear{2020ApJ...904L...4R}). 

Belbruno et al. (\citeyear{2012AsBio..12..754B}) studied the exchange of planetesimals between stars in an open cluster and the implications on the lithopanspermia hypothesis. Because in this work we are studying the exchange of planetesimals at a Galactic level, it is of interest to understand what constraints the processes mentioned earlier (that can alter the physical properties of the interstellar objects during ejection, transit and entry) impose on the possibility of lithopanspermia across the Galaxy.

\section{Conclusion}
\label{conclusion}

An unsolved problem in planet formation theory is how cm-sized pebbles grow into km-sized planetesimals, as particles approaching a meter size grow inefficiently due to increased collisional energies; in addition, they have short inward drift timescales due to gas drag that limit significantly their lifetime in the disk and their opportunity to grow to sizes unaffected by gas drag. Several mechanisms have been proposed to alleviate this meter-sized barrier. In this study we address one that has recently been suggested on the wake of the discovery of the first interstellar interlopers: the seeding of the star- and planet-forming environments with interstellar objects captured from the interstellar medium. These trapped objects could be large enough to experience a longer inward drift timescale and a higher mid-plane settling speed, favoring their rapid growth into larger bodies via the direct accretion of the sub-cm sized dust grains in the protoplanetary disk. 

In this work, we estimate the total number of interstellar objects in the 50 cm--50 km size range expected to be trapped by the molecular cloud and later incorporated to the disk of each forming stellar system, and also those expected to be trapped by the protoplanetary disk after it formed. Our calculations consider the uncertainties in the size distribution of the interstellar objects, and different assumptions regarding their origin and their expected background density. We also consider the case of the trapping of interstellar planetesimals by stars inside an open cluster, where the relative velocities of the stars are small and the exchange of planetesimals between them can be more effective. 

We find that the number of trapped interstellar objects can be significant. For example, when assuming a background number density of 2$\cdot$10$^{15}$ pc$^{-3}$ (based on 1I/'Oumuamua's detection), a velocity dispersion of 30 km/s and an equilibrium size distribution, the number of interstellar objects captured by a molecular cloud and expected to be incorporated to each protoplanetary disk during its formation is
6$\cdot$10$^{8}$ (50 cm--5 m), 
2$\cdot$10$^{5}$  (5 m--50 m), 
6$\cdot$10$^{1}$  (50 m--500 m), 
2$\cdot$10$^{-2}$  (500 m--5 km). After the disk formed, the number of interstellar objects that it could capture from the interstellar medium during its lifetime is  
6$\cdot$10$^{11}$ (50 cm--5 m), 
2$\cdot$10$^{8}$ (5 m--50 m), 
6$\cdot$10$^{4}$ (50 m--500 m), 
2$\cdot$10$^{1}$ (500 m--5 km).
The latter estimate assumes a field environment. 
In an open cluster environment, if 1\% of the clusters stars have undergone planet formation, the number of interstellar objects that a neighboring protoplanetary disk in the cluster could capture during the disk lifetime is a factor of $\sim$600 larger than the field values quoted above (1\% of the values listed in Table 6). 
As mentioned, these estimates correspond to an equilibrium size distribution. In this study we consider a wide range of possible size distributions for interstellar objects and expected background density. The corresponding results show a wide range of possible values (listed in Tables 3--6, for the different environments considered).

In the future, we will be able to narrow down these estimates as the population of interstellar objects becomes better characterized, both observationally and via simulations, in particular its background number density, size, and velocity distributions. This will soon be possible when the Vera Rubin Observatory starts operating, as it is expected that it will detect a large number of interstellar objects (could be one per month). Simulations to understand their distribution (in phase space and size) should take into account the ejection process of the planetesimals from their parent system and how their long journey in the Galaxy affects their initial velocity, as the interstellar planetesimal population is subject to size-dependent dynamical heating. Future simulations of molecular clouds and star-forming environments (Pfalzner et al. \citeyear{2021arXiv210608580P}) and of planet formation should take into account the presence of a population of trapped interstellar objects. The latter simulations are necessary to estimate how many of these interstellar seeds would be necessary to account for the planets and small body populations in solar and extra-solar planetary systems and to assess if the trapped interstellar objects can play an important role.

\hfill \break
We thank the referee Evgeni Grishin for numerous insightful suggestions that have considerably improved this study. We thank David Jewitt, Tushar Mittal and Stu Weidenschilling for stimulating conversations on this subject.

\appendix

\section{Effect of the Gas Drag}
\label{appendixA}
Table A1 in Appendix A lists the Mach number, $M = u/c$, corresponding to each environment under consideration and for a given interstellar object incoming velocity. The sound speed is given by $c = \left(\gamma k T_{\rm env}/m_0\right)^{1/2}$, where $k$ is Boltzmann's constant and $\gamma$ is the adiabatic index, adopting a value of 7/5 corresponding to diatomic molecules. For a given interstellar object radius (in the range of 50 cm to 50 km), Table A1 also lists the corresponding Reynolds number, calculated following Adachi et al. (\citeyear{1976PThPh..56.1756A}),  $R_{\rm e} = {2\rho_{\rm env}{\rm ur_{\rm p}} \over \mu}$. Here, $\mu$ is the viscosity,  $\mu$ = ${\rho_{\rm env}\rm{c_ml} \over 3}$, $l$ is the mean free path of the gas molecules, $l = {1 \over n_{\rm env} \sigma}$, $\sigma$ is the gas molecule cross section ($\sigma$ = 2.6$\cdot$10$^{-15}$ cm$^2$ for H$_2$), and $c_{m}$ is the mean thermal velocity of the gas molecules.  Because the latter is given by $c_{m} = \left(8 k T_{\rm env}/\pi m_{o}\right)^{1/2} = \left(8/\gamma \pi\right)^{1/2}c$, we have that $R_{\rm e} = 4.4 M {r_{\rm p} \over l} = 4.4 M K^{-1}$, where $K$ is the Knudsen number, $K=l/r_{\rm p}$, also listed in Table A1. 

The gas drag coefficient in Equation \eqref{Cd} can be expressed as a function of $R_{\rm e}$ and $K$ by using $M =  {K R_{\rm e} \over 4.4}$, giving 
\begin{equation}
\begin{split}
C_{\rm D} \simeq {24 \over R_{\rm e}} \left(\left(1 + {5 R_{\rm e} \over 3(10+R_{\rm e})}\right)^{-1} + 1.25 K\right)^{-1} 
+ \left({(2-w)K R_{\rm e}  \over 7.04+K R_{\rm e}} + w\right) 
= {C_1(R_{\rm e},K) \over R_{\rm e}} + C_2(R_{\rm e},K), 
\end{split}
\label{Cd2}
\end{equation}
where $C_1(R_{\rm e},K)$ and $C_2(R_{\rm e},K)$ are given by
\begin{equation}
C_{\rm 1} = 24 \left(\left(1 + {5 R_{\rm e} \over 3(10+R_{\rm e})}\right)^{-1} + 1.25 K\right)^{-1} \\
\label{C1}
\end{equation}
and
\begin{equation}
C_{\rm 2} = {(2-w) K R_{\rm e}  \over 7.04+K R_{\rm e}} + w, 
\label{C2}
\end{equation}
and are slowly varying over the range of $R_{\rm e}$ and $K$ considered. {\bf Equations \eqref{C1} and \eqref{C2} will be used in Appendix B.} 

\subsection{Stokes number}
Table A1 also lists Stokes number, $Stk$, which is the ratio of the stopping time, $t_{\rm s}$, to a dynamical time, ${2 R_{\rm env} \over u}$ (the crossing time of the cloud), so that $Stk = {t_{\rm s} u \over 2 R_{\rm env}}$. The stopping time is $t_{\rm s} = {x_{\rm f} - x_{\rm i} \over u}$, where $x_{\rm f} - x_{\rm i}$ is the distance travelled until the object halts to a stop. This distance can be calculated as follows. Consider an interstellar object with an incoming velocity $u$. It experiments an acceleration due to gas drag given by Equation \eqref{aD} so its position as a function of velocity is given by 
\begin{equation}
x(v) = \int {v \over a(v)} {\rm d} v =  {8 \rho_{\rm p} r_{\rm p} \over 3 \rho_{\rm env}}\int {1 \over C_{\rm D}(R_{\rm e},K)} {{\rm d} v\over v}.\\
\label{x}
\end{equation}
Replacing ${{\rm d} v\over v}$ by ${{\rm dR_{\rm e}} \over R_{\rm e}}$, we have that the distance travelled until the object halts to a stop, i.e. from when it has a Reynolds number of $R_{\rm e,i}$ until $R_{\rm e,f}$ = 0, is given by
\begin{equation}
\begin{split}
x_{\rm f} - x_{\rm i} =  {8 \rho_{\rm p} r_{\rm p} \over 3 \rho_{\rm env}}\int\limits_{R_{\rm e,i}}^{R_{\rm e,f=0}} {1 \over C_{\rm D}(R_{\rm e},K)} {{\rm d} R_{\rm e}\over R_{\rm e}} \\
= {8 \rho_{\rm p} r_{\rm p} \over 3 \rho_{\rm env}}\int\limits_{R_{\rm e,i}}^{R_{\rm e,f=0}} {1 \over {C_1(R_{\rm e},K) \over R_{\rm e}} + C_2(R_{\rm e},K)} {{\rm d} R_{\rm e}\over R_{\rm e}} \\
\simeq  {8 \rho_{\rm p} r_{\rm p} \over 3 \rho_{\rm env}C_2(R_{\rm e,i},K)}\left[{\rm ln}(C_1(R_{\rm e},K) + C_2(R_{\rm e},K) R_{\rm e})\right]_{R_{\rm e,i}}^{R_{\rm e,f=0}} \\
=  {8 \rho_{\rm p} r_{\rm p} \over 3 \rho_{\rm env}C_2(R_{\rm e,i},K)}{\rm ln}\left({C_1(R_{\rm e,i},K) \over C_1(R_{\rm e,i},K) + C_2(R_{\rm e,i},K) R_{\rm e,i}}\right),
\end{split}
\label{DistancetoHalt}
\end{equation}
where the integral has been calculated using Equation \eqref{Cd2} and assuming that $C_1(R_{\rm e},K)$ and $C_2(R_{\rm e},K)$ are constant over the range of $R_{\rm e}$ and $K$ considered.

Using the expression above, the Stokes number is given by 
\begin{equation}
\begin{split}
Stk(R_{\rm e},K) = {t_{\rm s} u \over 2 R_{\rm env}}=  {x_{\rm f} - x_{\rm i} \over 2 R_{\rm env}} \\
\simeq  {4 \rho_{\rm p} r_{\rm p} \over 3 \rho_{\rm env}C_2(R_{\rm e,i},K)R_{\rm env}}{\rm ln}\left({C_1(R_{\rm e,i},K) \over C_1(R_{\rm e,i},K) + C_2(R_{\rm e,i},K) R_{\rm e,i}}\right)\\
=  -{4 \rho_{\rm p} m_o \over 3 \rho_{\rm env}^2C_2(R_{\rm e,i},K)R_{\rm env}K \sigma}{\rm ln}\left(1+{C_2(R_{\rm e,i},K) \over C_1(R_{\rm e,i},K)} R_{\rm e,i}\right). 
\end{split}
\label{StkRK}
\end{equation}
To express the Stokes number as a function of $R_{\rm e}$ and $K$ we have replaced $r_{\rm p}$ by ${l \over K} = {m_o\over K \rho_{\rm env} \sigma}$, where $m_{o}$ and $\sigma$ are the mass and cross section of the gas molecule, respectively (assuming H$_2$,  $m_{0}$ = 3.34$\cdot$10$^{-24}$ g and $\sigma$ = 2.6$\cdot$10$^{-15}$ cm$^2$).
 
\begin{figure}
\begin{center}
\includegraphics[width=9cm]{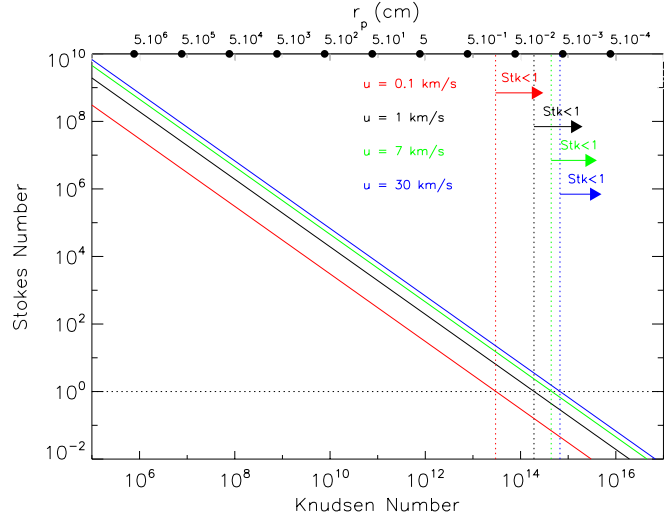}
\end{center}
\caption{Stokes number, $Stk$, as a function of the Knudsen number, $K$, calculated using Equation \eqref{StkMK},  assuming the conditions of the molecular cloud environment (see Table 1). The different colored lines correspond to different incoming velocities of the interstellar object with respect to the cloud. The top x-axis corresponds to the object's radius, given by $r_{p}=l/K$, where $l$ is the mean free path of the gas molecules. Interstellar objects with $Stk <<$ 1 are expected to be trapped by gas drag, while those with $Stk >>$ 1 are expected to escape. The dotted lines indicate the critical Knudsen number/object radius for which $Stk$ = 1}
\label{stokes_MC}
\end{figure}

\begin{figure}
\begin{center}
\includegraphics[width=9cm]{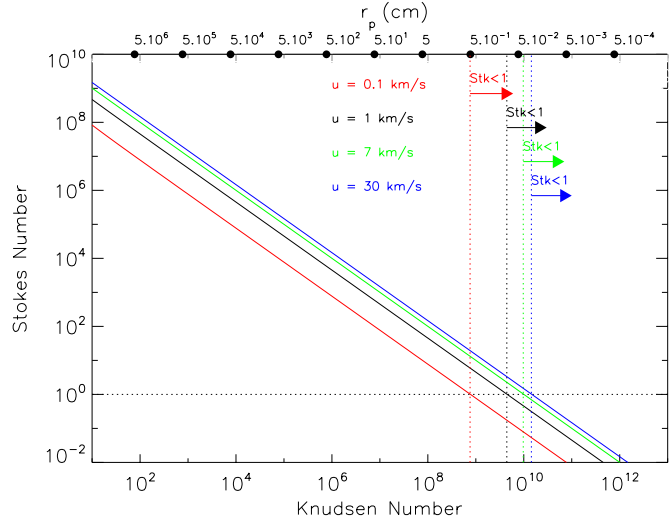}
\end{center}
\caption{Same as Figure \ref{stokes_MC} but for the stellar kernel environment (see Table 1).}
\label{stokes_kernel}
\end{figure}

\begin{figure}
\begin{center}
\includegraphics[width=9cm]{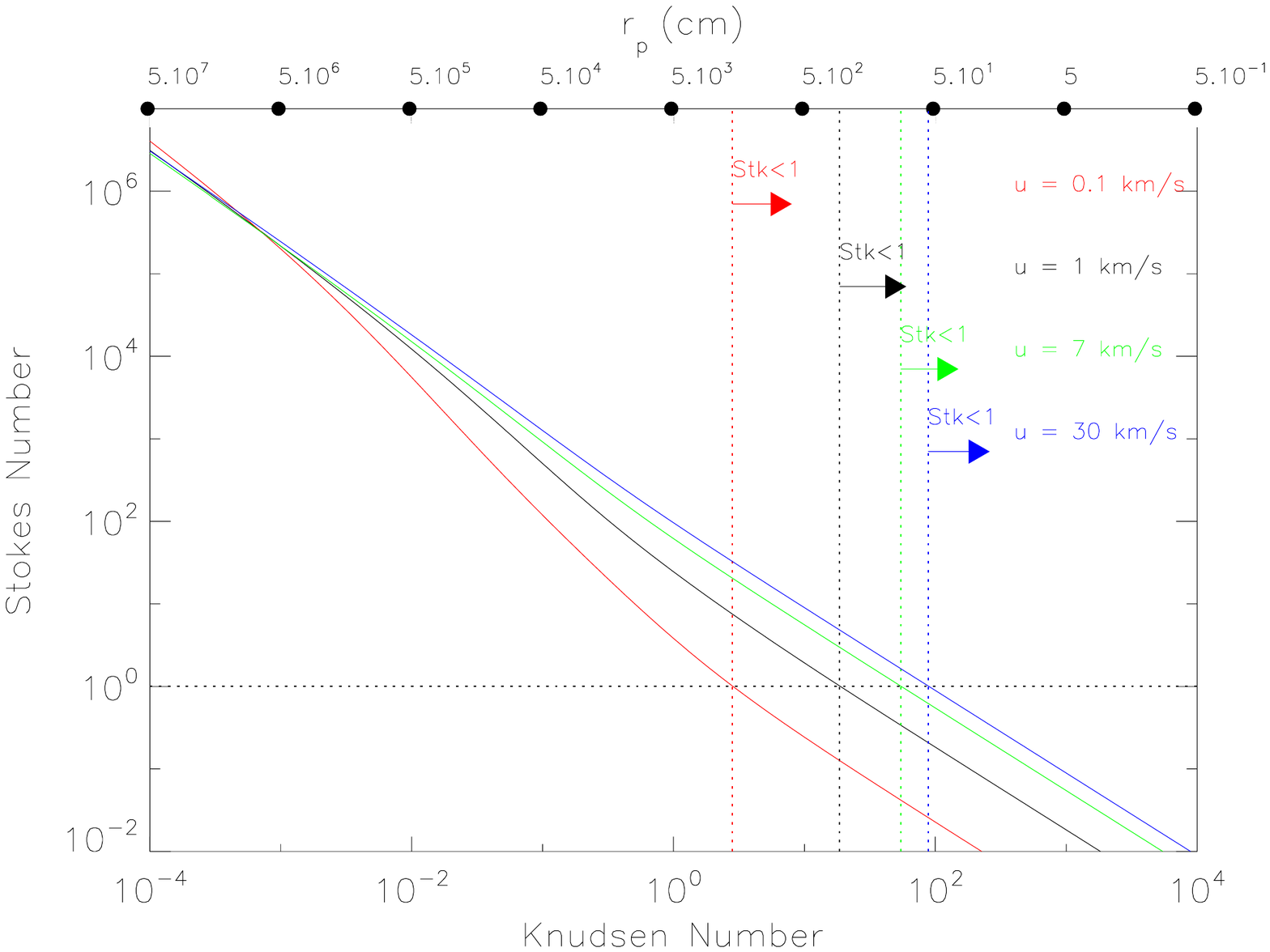}
\end{center}
\caption{Same as Figure \ref{stokes_MC} but for the MMSN-30au environment, based on Hayashi (\citeyear{1981PThPS..70...35H}; see Table 1).}
\label{stokes_MMSN30}
\end{figure}

Alternatively,  we can express Equation \eqref{StkRK} above as a function of $M$ and $K$ by using $R_{\rm e} = 4.4 M K^{-1}$ and putting $C_1$ and $C_2$ in Equations \eqref{C1} and \eqref{C2} as a function of $M$ and $K$. We then have
\begin{equation}
Stk = -{4 \rho_{\rm p} m_o \over 3 \rho_{\rm env}^2C_2(M,K)R_{\rm env}K \sigma}{\rm ln}\left(1+{C_2(M,K) \over C_1(M,K)}4.4 {M \over K}\right)\\
\label{StkMK}
\end{equation}
with  
\begin{equation}
C_{\rm 1} = 24 \left(\left(1 + {5 \over 3} {4.4 {M \over K} \over 10+4.4 {M \over K}}\right)^{-1} + 1.25 K\right)^{-1} \\
\label{C1MK}
\end{equation}
and
\begin{equation}
C_{\rm 2} = {(2-w) 4.4 M  \over 7.04+ 4.4 M} + w = {(2-w)   \over {1.6 \over M}+ 1} + w.
\label{C2MK}
\end{equation}
Figures \ref{stokes_MC}--\ref{stokes_MMSN30} show the Stokes number, $Stk$, as a function of the Knudsen number, $K$, calculated using Equations \eqref{StkMK}--\eqref{C2MK} for the different environments and incoming velocities under consideration. Table A2 shows the critical object radius, r$_{p}$ (cm), for which the Stokes number reaches the values of 1 and 0.1.  Interstellar objects with $Stk <<$ 1 are expected to be trapped by gas drag, while those with $Stk >>$ 1 are expected to escape.

\startlongtable 
\begin{deluxetable*}{lllllll}
\tablenum{A1}
\tablecaption{Effect of gas drag.\tablenotemark{a}}
\tablewidth{0pc}
\tablehead{
\colhead{} & 
\colhead{r$_p$ = 50 cm} & 
\colhead{r$_p$ = 5 m} & 
\colhead{r$_p$ = 50 m} & 
\colhead{r$_p$ = 500 m} & 
\colhead{r$_p$ = 5 km} &
\colhead{r$_p$ = 50 km}}
\startdata
\bf{Molecular cloud}\\
Knudsen number (K)			& K=7.69$\cdot$10$^{10}$	&  K=7.69$\cdot$10$^{9}$	& K=7.69$\cdot$10$^{8}$ 	& K=7.69$\cdot$10$^{7}$ 	& K=7.69$\cdot$10$^{6}$ 	& K=7.69$\cdot$10$^{5}$\\
\\
\bf{u = 0.1 km/s, M = 0.42} \\ 
Reynolds number (R$_{e}$)	& 2.38$\cdot$10$^{-11}$	& 2.38$\cdot$10$^{-10}$ 	& 2.38$\cdot$10$^{-9}$ 	& 2.38$\cdot$10$^{-8}$ 	& 2.38$\cdot$10$^{-7}$ 	& 2.38$\cdot$10$^{-6}$\\
Drag coefficient (C$_{D}$) 		& 11.08			        	& 11.08			        	& 11.08				& 11.08			 	& 11.08				& 11.08 \\
Stokes number 				& 3.96$\cdot$10$^{2}$  	& 3.96$\cdot$10$^{3}$  	& 3.96$\cdot$10$^{4}$  	& 3.96$\cdot$10$^{5}$  	& 3.96$\cdot$10$^{6}$  	&  3.96$\cdot$10$^{7}$  \\
\bf{u = 1 km/s, M = 4.16}	\\
Reynolds number (R$_{e}$)	& 2.38$\cdot$10$^{-10}$	& 2.38$\cdot$10$^{-9}$ 	& 2.38$\cdot$10$^{-8}$ 	& 2.38$\cdot$10$^{-7}$ 	& 2.38$\cdot$10$^{-6}$ 	& 2.38$\cdot$10$^{-5}$\\
Drag coefficient (C$_{D}$) 	& 2.60				& 2.60				& 2.60				& 2.60				& 2.60				& 2.60 \\
Stokes number 				& 2.51$\cdot$10$^{3}$  	& 2.51$\cdot$10$^{4}$  	& 2.51$\cdot$10$^{5}$  	& 2.51$\cdot$10$^{6}$  	& 2.51$\cdot$10$^{7}$  	&  2.51$\cdot$10$^{8}$  \\
\bf{u = 7 km/s, M = 29.10}	\\
Reynolds number (R$_{e}$)	& 1.66$\cdot$10$^{-9}$	& 1.66$\cdot$10$^{-8}$ 	& 1.66$\cdot$10$^{-7}$ 	& 1.66$\cdot$10$^{-6}$ 	& 1.66$\cdot$10$^{-5}$ 	& 1.66$\cdot$10$^{-4}$\\
Drag coefficient (C$_{D}$) 	& 2.07				& 2.07				& 2.07				& 2.07				& 2.07				& 2.07 \\
Stokes number 				& 5.88$\cdot$10$^{3}$  	& 5.88$\cdot$10$^{4}$  	& 5.88$\cdot$10$^{5}$  	& 5.88$\cdot$10$^{6}$  	& 5.88$\cdot$10$^{7}$  	& 5.88$\cdot$10$^{8}$   \\
\bf{u = 30 km/s, M = 124.74}	\\
Reynolds number (R$_{e}$)	& 7.13$\cdot$10$^{-9}$	& 7.13$\cdot$10$^{-8}$ 	& 7.13$\cdot$10$^{-7}$ 	& 7.13$\cdot$10$^{-6}$ 	& 7.13$\cdot$10$^{-5}$ 	& 7.13$\cdot$10$^{-4}$\\
Drag coefficient (C$_{D}$) 	& 2.01				& 2.01				& 2.01				& 2.01				& 2.01				& 2.01 \\
Stokes number 				& 8.80$\cdot$10$^{3}$  	& 8.80$\cdot$10$^{4}$  	& 8.80$\cdot$10$^{5}$  	& 8.80$\cdot$10$^{6}$  	& 8.80$\cdot$10$^{7}$  	& 8.80$\cdot$10$^{8}$   \\
\\
\hline
\\
\bf{Pre-stellar core kernel}\\
Knudsen number (K)			& K=7.69$\cdot$10$^{6}$	&  K=7.69$\cdot$10$^{5}$	& K=7.69$\cdot$10$^{4}$ 	& K=7.69$\cdot$10$^{3}$ 	& K=7.69$\cdot$10$^{2}$ 	& K=7.69$\cdot$10$^{1}$\\
\\
\bf{u = 0.1 km/s, M = 0.52} \\ 
Reynolds number (R$_{e}$)	& 2.95$\cdot$10$^{-7}$	& 2.95$\cdot$10$^{-6}$ 	& 2.95$\cdot$10$^{-5}$ 	& 2.95$\cdot$10$^{-4}$ 	& 2.95$\cdot$10$^{-3}$ 	& 2.95$\cdot$10$^{-2}$\\
Drag coefficient (C$_{D}$) 		& 9.15				& 9.15				& 9.15				& 9.15				& 9.15				& 9.07 \\
Stokes number 				& 1.01$\cdot$10$^{2}$  	& 1.01$\cdot$10$^{3}$  	& 1.01$\cdot$10$^{4}$  	& 1.01$\cdot$10$^{5}$  	& 1.01$\cdot$10$^{6}$  	& 1.01$\cdot$10$^{7}$  \\
\bf{u = 1 km/s, M = 5.16}	\\
Reynolds number (R$_{e}$)	& 2.95$\cdot$10$^{-6}$	& 2.95$\cdot$10$^{-5}$	& 2.95$\cdot$10$^{-4}$ 	& 2.95$\cdot$10$^{-3}$ 	& 2.95$\cdot$10$^{-2}$	& 2.95$\cdot$10$^{-1}$\\
Drag coefficient (C$_{D}$) 	& 2.46				& 2.46				& 2.46				& 2.46				& 2.46				& 2.46 \\
Stokes number 				& 5.87$\cdot$10$^{2}$  	& 5.87$\cdot$10$^{3}$  	& 5.87$\cdot$10$^{4}$  	& 5.87$\cdot$10$^{5}$  	& 5.88$\cdot$10$^{6}$  	& 5.91$\cdot$10$^{7}$  \\
\bf{u = 7 km/s, M = 36.1}	\\
Reynolds number (R$_{e}$)	& 2.06$\cdot$10$^{-5}$	& 2.06$\cdot$10$^{-4}$ 	& 2.06$\cdot$10$^{-3}$ 	& 2.06$\cdot$10$^{-2}$ 	& 2.06$\cdot$10$^{-1}$ 	& 2.06\\
Drag coefficient (C$_{D}$) 	& 2.05				& 2.05				& 2.05				& 2.05				& 2.05				& 2.05 \\
Stokes number 				& 1.30$\cdot$10$^{3}$  	& 1.30$\cdot$10$^{4}$  	& 1.30$\cdot$10$^{5}$  	& 1.30$\cdot$10$^{6}$  	& 1.30$\cdot$10$^{7}$  	& 1.31$\cdot$10$^{8}$  \\
\bf{u = 30 km/s, M = 154.72}	\\
Reynolds number (R$_{e}$)	& 8.85$\cdot$10$^{-5}$	& 8.85$\cdot$10$^{-4}$ 	& 8.85$\cdot$10$^{-3}$ 	& 8.85$\cdot$10$^{-2}$ 	& 8.85$\cdot$10$^{-1}$ 	& 8.85\\
Drag coefficient (C$_{D}$) 	& 2.01				& 2.01				& 2.01				& 2.01				& 2.01				& 2.01 \\
Stokes number 				& 1.91$\cdot$10$^{3}$  	& 1.91$\cdot$10$^{4}$  	& 1.91$\cdot$10$^{5}$  	& 1.91$\cdot$10$^{6}$  	& 1.91$\cdot$10$^{7}$  	& 1.92$\cdot$10$^{8}$  \\
\\
\hline
\\
\bf{MMSN-30au}\\
Knudsen number (K)			& K=9.62$\cdot$10$^{1}$	&  K=9.62		& K=9.62$\cdot$10$^{-1}$ 	& K=9.62$\cdot$10$^{-2}$ 	& K=9.62$\cdot$10$^{-3}$ 	& K=9.62$\cdot$10$^{-4}$\\
\\
\bf{u = 0.1 km/s, M = 0.19} \\ 
Reynolds number (R$_{e}$)	& 8.51$\cdot$10$^{-3}$	& 8.51$\cdot$10$^{-2}$	& 8.51$\cdot$10$^{-1}$ 	& 8.51 	& 8.51$\cdot$10$^{1}$ 	& 8.51$\cdot$10$^{2}$\\
Drag coefficient (C$_{D}$) 		& 23.27				& 22.76				& 13.89				& 4.66				& 1.25				& 0.64 \\
Stokes number 				& 2.36$\cdot$10$^{-2}$  (**)	& 2.53$\cdot$10$^{-1}$  (*)	& 4.03 	&  1.27$\cdot$10$^{2}$  		& 5.93$\cdot$10$^{3}$  		& 2.11$\cdot$10$^{5}$  \\
\bf{u = 1 km/s, M = 1.86}	\\
Reynolds number (R$_{e}$)	& 8.51$\cdot$10$^{-2}$	& 8.51$\cdot$10$^{-1}$	& 8.51 	& 8.51$\cdot$10$^{1}$ 	& 8.51$\cdot$10$^{2}$	& 8.51$\cdot$10$^{3}$\\
Drag coefficient (C$_{D}$) 	& 3.58				& 3.44				& 2.85				& 1.80				& 1.33				& 1.27 \\
Stokes number 				& 0.19$\cdot$10$^{-1}$  (*)	& 2.01  	& 2.57$\cdot$10$^{1}$  	&  5.31$\cdot$10$^{2}$  		& 1.29$\cdot$10$^{4}$  		& 2.26$\cdot$10$^{5}$  \\
\bf{u = 7 km/s, M = 13.01}	\\
Reynolds number (R$_{e}$)	& 5.96$\cdot$10$^{-1}$	& 5.96 	& 5.96$\cdot$10$^{1}$ 	& 5.96$\cdot$10$^{2}$ 	& 5.96$\cdot$10$^{3}$ 	& 5.96$\cdot$10$^{4}$\\
Drag coefficient (C$_{D}$) 	& 2.16				& 2.14				& 2.07				& 1.91				& 1.83				& 1.83 \\
Stokes number 				& 0.57$\cdot$10$^{-1}$  (*)	& 5.80  	& 6.45$\cdot$10$^{1}$  	&  9.63$\cdot$10$^{2}$  		& 1.58$\cdot$10$^{4}$  		& 2.27$\cdot$10$^{5}$  \\
\bf{u = 30 km/s, M = 55.78}	\\
Reynolds number (R$_{e}$)	& 2.55	& 2.55$\cdot$10$^{2}$ 	& 2.55$\cdot$10$^{3}$ 	& 2.55$\cdot$10$^{4}$ 	& 2.55$\cdot$10$^{5}$ 	& 2.55$\cdot$10$^{6}$\\
Drag coefficient (C$_{D}$) 	& 2.03				& 2.01				& 1.97				& 1.96				& 1.95				& 1.95\\
Stokes number 				& 9.93$\cdot$10$^{-1}$ (*) 	& 9.36  	& 1.00$\cdot$10$^{2}$  	&  1.32$\cdot$10$^{3}$  		& 1.90$\cdot$10$^{4}$  		& 2.56$\cdot$10$^{5}$  \\
\\
\hline
\\
\bf{DISK-30au}\\
Knudsen number (K)			& K=1.54$\cdot$10$^{3}$	&  K=1.54$\cdot$10$^{2}$		& K=1.54$\cdot$10$^{1}$	& K=1.54 	& K=1.54$\cdot$10$^{-1}$ 	& K=1.54$\cdot$10$^{-2}$\\
\\
\bf{u = 0.1 km/s, M = 0.19} \\ 
Reynolds number (R$_{e}$)	& 5.31$\cdot$10$^{-4}$	& 5.31$\cdot$10$^{-3}$	& 5.31$\cdot$10$^{-2}$ 	& 5.31$\cdot$10$^{-1}$ 	& 5.31	& 5.31$\cdot$10$^{1}$\\
Drag coefficient (C$_{D}$) 		& 23.44				& 23.34				& 22.36				& 16.17				& 6.01				& 1.61 \\
Stokes number 				& 3.74$\cdot$10$^{-1}$ (*) 	& 3.76	& 3.93$\cdot$10$^{1}$ 	&  5.51$\cdot$10$^{2}$  		& 1.55$\cdot$10$^{4}$  		& 6.85$\cdot$10$^{5}$  \\
\bf{u = 1 km/s, M = 1.86}	\\
Reynolds number (R$_{e}$)	& 5.31$\cdot$10$^{-3}$	& 5.31$\cdot$10$^{-2}$	& 5.32$\cdot$10$^{-1}$ 	& 5.32 	& 5.32$\cdot$10$^{1}$	& 5.32$\cdot$10$^{2}$\\
Drag coefficient (C$_{D}$) 	& 3.60				& 3.59				& 3.49				& 3.02				& 2.00				& 1.37 \\
Stokes number 				& 3.03  	& 3.05$\cdot$10$^{1}$	& 3.15$\cdot$10$^{2}$ 	&  3.80$\cdot$10$^{3}$  		& 7.01$\cdot$10$^{4}$  		& 1.76$\cdot$10$^{6}$  \\
\bf{u = 7 km/s, M = 13.01}	\\
Reynolds number (R$_{e}$)	& 3.72$\cdot$10$^{-2}$	& 3.72$\cdot$10$^{-1}$	& 3.72 	& 3.72$\cdot$10$^{1}$ 	& 3.72$\cdot$10$^{2}$ 	& 3.72$\cdot$10$^{3}$\\
Drag coefficient (C$_{D}$) 	& 2.16				& 2.16				& 2.15				& 2.10				& 1.94				& 1.84\\
Stokes number 				& 9.08 	& 9.10$\cdot$10$^{1}$  	& 9.22$\cdot$10$^{2}$  	&  9.93$\cdot$10$^{3}$  		& 1.39$\cdot$10$^{5}$  		& 2.30$\cdot$10$^{6}$  \\
\bf{u = 30 km/s, M = 55.78}	\\
Reynolds number (R$_{e}$)	& 1.60$\cdot$10$^{-1}$	& 1.60 	& 1.60$\cdot$10$^{1}$	& 1.60$\cdot$10$^{2}$	& 1.60$\cdot$10$^{3}$ 	& 1.60$\cdot$10$^{4}$\\
Drag coefficient (C$_{D}$) 	& 2.03				& 2.03				& 2.03				& 2.02				& 1.98				& 1.96\\
Stokes number 				& 1.48$\cdot$10$^{1}$ 	& 1.48$\cdot$10$^{2}$  	& 1.49$\cdot$10$^{3}$  	&  1.56$\cdot$10$^{4}$  		& 1.96$\cdot$10$^{5}$ 		& 2.84$\cdot$10$^{6}$  \\
\enddata
\tablenotetext{a}{(*) indicates objects with $Stk < 1$. (**) indicates objects with $Stk < 0.1$.Particles with  $Stk << 1$ are expected to get trapped.}
\end{deluxetable*}

\startlongtable 
\begin{deluxetable*}{llll}
\tablenum{A2}
\tablecaption{Critical object radius, r$_{p}$ (cm), for gas trapping}
\tablewidth{0pc}
\tablehead{
\colhead{} &
\colhead{Molecular} &
\colhead{Pre-stellar} &
\colhead{MMSN-30au}
\\
\colhead{} &
\colhead{cloud} &
\colhead{kernel} &
\colhead{}
\\
\colhead{} &
\colhead{(Stk=1)} &
\colhead{(Stk=1)} &
\colhead{(Stk=1)}
}
\startdata
u = 0.1 	& 0.1			& 0.5			& 2$\cdot$10$^{3}$ 	\\
u = 1		& 0.02		& 0.09		& 3$\cdot$10$^{2}$	\\
u = 7		& 0.009		& 0.04		& 9$\cdot$10$^{1}$	\\
u = 30	& 0.006		& 0.03		& 6$\cdot$10$^{1}$	\\
\enddata
\end{deluxetable*}

For the molecular cloud and stellar kernel environments, characteristic of the earliest stages of star and planet formation, all particles in the 50 cm--50 km radius range have $Stk >> 1$, even when considering very low incoming velocities. This indicates that gas drag is not an efficient mechanism for the trapping of meter-size particles in those environments. 

Another way to see this is to compare the column density of the incoming interstellar object, ${m \over \pi r^2}$, to that of the gas it will swept as it moves through the cloud,  $2R_{\rm env}\rho_{\rm env}$. We find that for interstellar objects with a radius in the range $r$ = 50 cm--50 km, we have that ${m \over \pi r^2} \sim 7\cdot10^{1}$--$7\cdot10^{6}$ g $\cdot$ cm$^{-2}$. This compares to $2R_{\rm env}\rho_{\rm env} \sim 3\cdot10^{-2}$ g$\cdot$cm$^{-2}$, for the molecular cloud environment, and $2R_{\rm env}\rho_{\rm env} \sim 10^{-1}$ g$\cdot$cm$^{-2}$, for the pre-stellar core kernel environment (assuming in both cases a molecular hydrogen composition). In both cases, the column density of the gas is significantly lower that that of the object and the particle has too much momentum to be trapped by gas drag. 

At a later stage of the star and planet formation process, the characteristic gas density and temperature are higher. Tables 2 and 3 and Figure \ref{stokes_MMSN30} show that, when considering the densities and temperatures characteristic of the mid-plane of the minimum mass solar nebula at 30 au (labeled as MMSN-30au in Tables 1 and 2), interstellar planetesimals with radius $<$ 50 cm would have $Stk < 1$ for all the incoming velocities considered, but $Stk > 1$ in the meter-sized range, except at the slowest incoming velocities.

Here, we are approximating this environments as spherical and having a uniform density, rather than the structured protoplanetary disks they really are. A detailed study of the trapping of planetesimals by gas drag during the protoplanetary disk phase that takes into account their structure is given by Grishin et al. (\citeyear{2019MNRAS.487.3324G}).

\section{Capture Conditions}
\label{appendixB}
Grishin et al. (\citeyear{2019MNRAS.487.3324G}) estimated the capture conditions by calculating the energy dissipated by the particle during its passage through the environment. Capture takes place when the energy loss is larger than the particle's kinetic energy at infinity, $\Delta E > E_0$, with $E_0 = {1\over2} m_p u_0^2$. In their calculation of the energy dissipated, they assumed that the acceleration that the particle experiments due to gas drag is given by Equation \eqref{aD} and adopted a constant value for the drag coefficient. However, in our calculations in Section \ref{appendixA}, we are taking into account the velocity dependency of the gas drag coefficient, given by Equation \eqref{Cd2}, so instead of Equation \eqref{aD}, the gas drag the particle experiments would be given by  
\begin{equation}
{{\rm d}u \over {\rm d}t} = -{\pi r_p^2 \over 2 m_p}\rho_{env}\left(C_1(u)u {K c \over 4.4} + C_2(u)u^2\right), 
\label{capcon2}    
\end{equation}
where $c$ is the sound speed and C$_1$(u) and C$_2$(u) are given by Equations \ref{C1} and \ref{C2}. 
Defining $\bar{C_1}(u,K) = C_1 (u){Kc\over4.4}$ and $\Sigma_p = {m_p \over \pi r_p^2}$, 

\begin{equation}
{{\rm d}u \over {\rm d}t} = -{1 \over 2 \Sigma_p}\rho_{env}\left(\bar{C_1}(u,K)u  + C_2(u)u^2\right). 
\label{capcon2}    
\end{equation}

We then have, 
\begin{equation}
{1 \over u} { {\rm d}u\over {\rm d}t} = { {\rm d}u\over {\rm d}x} = -{1 \over 2 \Sigma_p}\rho_{env}(x)\left(\bar{C_1}(u,K)  + C_2(u)u\right). 
\label{capcon2}    
\end{equation}

Defining
\begin{equation}
\begin{split}
y \equiv \bar{C_1}(u,K) \int {1 \over 2 \Sigma_p}\rho_{env}(x){\rm d}x\\
\end{split}
\end{equation}
and
\begin{equation}
\begin{split}
\eta \equiv {C_2(u) \over \bar{C_1}(u,K)}, 
\end{split}
\end{equation}
we have that
\begin{equation}
\begin{split}
{\rm d}y =  {\bar{C_1}(u,K) \over 2 \Sigma_p}\rho_{env}(x){\rm d}x,\\
\end{split}
\end{equation}
\begin{equation}
\begin{split}
{{\rm d}u \over {\rm d}y} = - \left(1 + {C_2(u)u \over \bar{C_1}(u,K)}\right) = - \left(1 + \eta\right), \\
\end{split}
\end{equation}
and 
\begin{equation}
\begin{split}
-{\rm d}y = {{\rm d}u \over 1 + \eta u}. 
\end{split}
\label{capcon2}    
\end{equation}
Solving for the velocity, $u$, we get
\begin{equation}
u = \eta^{-1}\left[A e^{-\eta\int{\rm d}y} -1 \right], 
\end{equation}
where  $A \equiv e^{C\eta}$ and $C$ is a constant of integration. This leads to
\begin{equation}
u = {\bar{C_1}(u,K) \over C_2(u)}\left[A e^{- C_2(u)\int\limits_{0}^{x} {\rho_{\rm env}(x)  \over  2 \Sigma_p}{\rm d}x} -1 \right].
\end{equation}
The initial condition $x=0$, $u=u_0$ gives, 
\begin{equation}
1+  {C_2(u_0) \over \bar{C_1}(u_0,K)}u_0 = A. 
\label{capcon1}
\end{equation}
so that 
\begin{equation}
\begin{split}
1+  {C_2(u) \over \bar{C_1}(u,K)}u  = 
\left(1+  {C_2(u_0) \over \bar{C_1}(u_0,K)}u_0\right)e^{{- C_2(u) \over 2 \Sigma_p} \int\limits_{0}^{x}\rho_{\rm env}(x)  {\rm d}x}.
\label{capcon2}
\end{split}
\end{equation}
For small ${C_2(u) \over 2 \Sigma_p} \int\limits_{0}^{x}\rho_{\rm env}(x)  {\rm d}x$, we get that
\begin{equation}
\begin{split}
u = {C_2(u_0) \over C_1(u_0)} {C_1(u) \over C_2(u)} u_0 
- {1 \over 2 \Sigma_p} \left(\bar{C_1}(u_0,K) + { C_2(u_0) C_1(u) \over C_1(u_0)}\right) \int\limits_{0}^{x}\rho_{\rm env}(x)  {\rm d}x.  
\end{split}    
\end{equation}
Generally, we have 
\begin{equation}
\begin{split}
{\bar{C_1}(u,K) +   C_2(u) u \over \bar{C_1}(u_0,K) +   C_2(u_0) u_0} 
= e^{{- C_2(u) \over 2 \Sigma_p} \int\limits_{0}^{x}\rho_{\rm env}(x)  {\rm d}x}, 
\label{capcon2}
\end{split}
\end{equation}

\begin{equation}
\begin{split}
- {2 \Sigma_p \over C_2(u)}{\rm ln}\left({\bar{C_1}(u,K) +  C_2(u) u \over \bar{C_1}(u_0,K) +   C_2(u_0) u_0}\right) =  \int\limits_{0}^{x}\rho_{\rm env}(x)  {\rm d}x. 
\end{split}
\label{capcon3}
\end{equation}
Here we can assume that $\bar{C_1}$ and $C_2$ are slowly varying, so that $\bar{C_1}(u,K) \simeq \bar{C_1}(u_0,K)$ and $C_2(u) \simeq C_2(u_0)$, that can be evaluated using Equations \eqref{C1} and \eqref{C2}. The resulting expression, 
\begin{equation}
\begin{split}
- {2 \Sigma_p \over C_2(u_0)}{\rm ln}\left({\bar{C_1}(u_0,K) +  C_2(u_0) u \over \bar{C_1}(u_0,K) +  C_2(u_0) u_0}\right) = 
  \int\limits_{0}^{x}\rho_{\rm env}(x)  {\rm d}x. 
\end{split}
\label{capcon3}
\end{equation}
can be solved numerically to calculate the velocity of the particle $u$ as a function of the mass it sweeps up as it traverses the environment. Capture takes place when the energy dissipated is larger than the particle's kinetic energy at infinity.

We now look at the case of a spherical cloud. 

The energy dissipated by the particle as it crosses a spherical cloud of uniform density, $\rho_{\rm env}$, is 
\begin{equation}
\begin{split}
\Delta E  = \int {F_{\rm D}} {\rm d}x = -{1\over2} \int {C_{\rm D}(R_{\rm e},K)\pi r_{\rm p}^{2} \rho_{env} u^{2}(x)} {\rm d}x \\
= -{\pi r_p^2 \over 2}\rho_{env}\int\left(\bar{C_1}(u_0,K)u + C_2(u)u^2\right)u{\rm d}t.
\label{deltaE}
\end{split}
\end{equation}
where the integral is along the perturbed orbit, 
and, as we indicated above,  $\bar{C_1}$ and $C_2$ are slowly varying. The two limiting cases are when the particle goes through the center of the cloud and when the particle grazes the cloud in an almost parabolic orbit.

\subsection{Particle crossing the center of the cloud}
Because a cloud of constant density has an harmonic oscillator potential well, the position of the particle is $x(t) = - a {\rm cos}(wt)$, where the harmonic oscillator period and frequency are
$T = \sqrt{{3 \pi}\over{G \rho_{\rm env}}}= {{2\pi}\over w}$ and $\dot{x}(t) = a w {\rm sin}(wt)$. The velocity of the particle is therefore $u(t)=u_{\rm 0}+u_{\rm esc} {\rm sin} (wt)$. We now calculate how much energy is dissipated in half a cycle to see if the particle will come back (i.e. if the particle is trapped by the environment). We have that
\begin{equation}
\begin{split}
\Delta E  = -{\pi r_p^2 \over 2}\rho_{env}\int\left[\bar{C_1}(u,K)u  + C_2(u)u^2\right]u{\rm d}t \\
= -{\pi r_p^2 \over 2}\rho_{env}
\int_0^{\pi\over\omega}[\bar{C_1}(u,K)(u_{\rm 0}+u_{\rm esc} {\rm sin} (wt))^2  
+ C_2(u)(u_{\rm 0} + u_{\rm esc} {\rm sin} (wt))^3] {\rm d}t.
\label{deltaE}
\end{split}
\end{equation}
As above, we can assume that $C_1$ and $C_2$ are slowly varying, so that $\bar{C_1}(u,K) \simeq \bar{C_1}(u_0,K)$ and $C_2(u) \simeq C_2(u_0)$, leading to
\begin{equation}
\begin{split}
\Delta E  = -{\pi r_p^2 \over 2}\rho_{env}
[{\pi \over \omega} (\bar{C_1}(u_0,K)u_{\rm 0}^2  + C_2(u_0)u_{\rm 0}^3 +
{\bar{C_1}(u_0,K) \over \omega} (4u_{\rm 0} u_{\rm esc} + {\pi\over2} u_{\rm esc}^2)+
{C_2(u_0) \over \omega} ({3\pi \over 2}u_{\rm 0} u_{\rm esc}^2 + 
6u_{\rm 0}^2 u_{\rm esc} + {4 \over 3} u_{\rm esc}^3)] 
\label{deltaE}
\end{split}
\end{equation}
and
\begin{equation}
\begin{split}
{\Delta E \over E_0} = 
-{ \rho_{env} \over \Sigma_p  \omega}
[\pi(\bar{C_1}(u_0,K) + C_2(u_0)u_{\rm 0})
+ \bar{C_1}(u_0,K) (4{u_{\rm esc} \over u_{\rm 0}} + {\pi\over2} {u_{\rm esc}^2\over u_0^2})+
C_2(u_0) u_0 ({3\pi \over 2}{u_{\rm esc}^2 \over u_{\rm 0}^2} + 
6{u_{\rm esc} \over u_{\rm 0}} + {4 \over 3} {u_{\rm esc}^3\over u_0^3})]. 
\label{deltaE}
\end{split}
\end{equation}
Defining $x={u_{\rm esc}\over u_0}$, we have that
\begin{equation}
\begin{split}
{\Delta E \over E_0} = 
-{\rho_{env} \over \Sigma_p  \omega}
[\pi\bar{C_1}(u_0,K) + C_2(u_0)u_{\rm 0})
+ \bar{C_1}(u_0,K) (4x + {\pi\over2} x^2)+
C_2(u_0) u_0 ({3\pi \over 2}x^2 + 
6x + {4 \over 3} x^3)]. 
\label{deltaE}
\end{split}
\end{equation}

To take into account the effect of the gravitational focusing outside the cloud, in the expression for $\Delta E$ above, we replace $u_0$ by $u_0 \sqrt{1+{u_{\rm esc}^2\over u_0^2}}$, leading to 
\begin{equation}
\begin{split}
\Delta E  = -{\pi r_p^2 \over 2}\rho_{env}
[{\pi \over \omega} (\bar{C_1}(u_0,K)u_{\rm 0}^2 
({1+{u_{\rm esc}^2\over u_0^2}})  
+ C_2(u_0)u_{\rm 0}^3 ({1+{u_{\rm esc}^2\over u_0^2}})^{3 \over 2}  
+{\bar{C_1}(u_0,K) \over \omega} (4u_{\rm 0} u_{\rm esc} ({1+{u_{\rm esc}^2\over u_0^2}})^{1 \over 2} + {\pi\over2} u_{\rm esc}^2)\\
+ {C_2(u_0) \over \omega} ({3\pi \over 2}u_{\rm 0} u_{\rm esc}^2 ({1+{u_{\rm esc}^2\over u_0^2}})^{1 \over 2} 
+ 6u_{\rm 0}^2 u_{\rm esc}({1+{u_{\rm esc}^2\over u_0^2}}) + {4 \over 3} u_{\rm esc}^3)]. 
\label{deltaE}
\end{split}
\end{equation}
Defining $x={u_{\rm esc}\over u_0}$ and replacing $\omega = \sqrt{4 \pi G \rho_{env}\over 3}$, we get an expression for ${\Delta E \over E_0}$ as a function of the environment parameters ($\rho_{\rm env}$, $R_{\rm env}$, $T_{\rm env}$),  particle parameters ($r_p$, $m_p$) and velocity at infinity, $u_0$,
\begin{equation}
\begin{split}
{\Delta E \over E_0} =
-{\rho_{env} \over \Sigma_p}\sqrt{3 \over 4 \pi G \rho_{env}}
[\pi(\bar{C_1}(u_0,K)) (1+x^2) 
+ C_2(u_0)u_{\rm 0})(1+x^2)^{3 \over 2}
+ \bar{C_1}(u_0,K) (4x (1+x^2)^{1 \over 2}+ {\pi\over2} x^2)\\
+ C_2(u_0) u_0 ({3\pi \over 2}x^2 (1+x^2)^{1 \over 2}+ 
6x(1+x^2) + {4 \over 3} x^3)].
\label{deltaE}
\end{split}
\end{equation}

Trapping takes place when the final energy is less than the binding energy to the cloud, 
\begin{equation}
\begin{split}
{E_f \over E_{BE}} = {E_0 + \Delta E \over E_{BE}} = {1 + {\Delta E \over E_0} \over {E_{BE}\over E_0}} = {{1 + {\Delta E \over E_0} \over x^2} < 1}, 
\label{TC}
\end{split}
\end{equation}
so that $-{\Delta E \over E_0} > 1 - x^2$. 
This limiting case where the particle crosses through the center of the cloud corresponds to maximum dissipation. 

\subsection{Particle grazing the cloud}

We now look at the case of lowest dissipation, where the particle enters the cloud in an almost parabolic orbit and gets trapped in a circular orbit or radius $a$. In this case, the energy dissipated is a function of $a$,
\begin{equation}
\begin{split}
\Delta E  = -{\pi r_p^2 \over 2}\rho_{env}\int\left[\bar{C_1}(u_{\rm cir})u_{\rm cir} + C_2(u_{\rm cir})u_{\rm cir}^2\right]{\rm d}x 
= - \psi a \pi^2 r_p^2 \rho_{env}\left[\bar{C_1}(u_{\rm cir})u_{\rm cir} + C_2(u_{\rm cir})u_{\rm cir}^2\right],   
\label{deltaEgrazing}
\end{split}
\end{equation}
where $\psi$ is of order 1 and depends on the exact nature of the incoming orbit. When the orbit is just trapped, it moves on an ellipse with orbital period $T = \sqrt{2\pi \over G\rho}$, independent of the position in the cloud. The motion is a circular orbit with an epicyclic motion in the square of the radial coordinate (it is not linear in the radial coordinate; see Harter Soft Web Resources for Physics, Chapter 5.2). Once the orbit is just trapped as an ellipse in the cloud, it will continue to dissipate. Since the highest velocity occurs at the minor axis, the orbit will circularized rather than become more elongated. For the geometrical image see Figure 5.2.2 and 5.2.1 in Harter (Soft Web Resources for Physics). 

In the geometric regime, when the particle is grazing the cloud, the impact parameter is $b = a = R_{\rm env}$ and we have that $u_0 = u_{\rm cir} = b\sqrt{{4\over3} G \rho_{\rm env}}$ and $E_0 = {1\over 2}m_p u_{\rm cir}^2 = {1\over 2}m_p b^2 {4\over3} G \rho_{\rm env}$. There are two limiting cases for $u_0$. The first one is when the parabolic orbit is just trapped to a circular orbit at the edge of the cloud, $u_0 = u_{\rm cir}$. The second limiting case is for small impact parameters, when the orbit oscillates radially across the cloud; in this case, the circular velocity squared is of the order of the depth of the potential well and therefore $u$ averaged over the orbit is of the order of $\chi v_{\rm cir}$, where $\chi$ is of order unity and depends on the exact nature of the orbit. 
\begin{equation}
\begin{split}
{\Delta E \over E_0} = - \psi {3 \over 2 \Sigma_p G b}
[\bar{C_1}(\chi u_{\rm cir}) \chi u_{\rm cir} + C_2(\chi u_{\rm cir}) \chi^2 u_{\rm cir}^2]. 
\label{deltaEE0grazing_gr}
\end{split}
\end{equation}
Trapping takes place when $-{\Delta E \over E_0} > 1 - x^2$, where $x = {u_{\rm esc} \over u_{0}}$ and $u_{\rm esc}$ is the escape velocity from the cloud. 

In the gravitational focusing regime, $a = q$, where $q$ is the distance of closest approach. In the parabolic approximation, $a = q \approx {b^2 u_0^2 \over 2 GM}$. Because of conservation of angular momentum, we have that $b^2 u_0^2 = q^2 u_{\rm cir}^2(a)$, where $b$ is the impact parameter and $u_{\rm cir} = \sqrt{2}{GM\over b u_0}$.
The initial energy can then be given as a function of the impact parameter, $E_0 = {1\over 2}m_p u_{\rm cir}^2 = m_p \left({GM \over b u_0}\right)^2$. We then have,
\begin{equation}
\begin{split}
{\Delta E \over E_0} = -\psi {\rho_{\rm env} b^4 u_0^4 \over 2 \Sigma_p G^3 M^3}
\left[\bar{C_1}(u_{\rm cir}) u_{\rm cir} + C_2(u_{\rm cir})u_{\rm cir}^2\right]. 
\label{deltaEE0grazing_gf}
\end{split}
\end{equation}
Following Equation \eqref{TC} above, trapping would take place when $-{\Delta E \over E_0} > 1 - x^2$, i.e. 
\begin{equation}
\begin{split}
\psi {\rho_{\rm env} b^4 u_0^4 \over 2 \Sigma_p G^3 M^3}
\left[\bar{C_1}(u_{\rm cir}) u_{\rm cir} + C_2(u_{\rm cir})u_{\rm cir}^2\right] > 1 - {u_{\rm esc}^2 \over u_{0}^2}.
\label{TCgrazing}
\end{split}
\end{equation}

To calculate the capture rates, we would need to use the critical capture condition, $-{\Delta E \over E_0} = 1 - x^2$ (i.e. Equation  \eqref{TCgrazing} with an equality), to get an expression for the critical impact parameter $b$ for capture, as a function of the environment parameters ($\rho_{\rm env}$, $R_{\rm env}$, $T_{\rm env}$, M),  particle parameters ($r_p$, $m_p$), and velocity at infinity, $u_0$.  However, the dependency of Equation \eqref{TCgrazing} on $u_{\rm cir}$ (which depends on $b$) does not allow to solve analytically for  $b$. This is why in Section \ref{capturerates} we used a simplified expression for the drag force, $F_{\rm D} = {1\over2 }C_{\rm D}(u)\pi r_{\rm p}^{2} \rho_{env} u^{2} = {3 C_{\rm D}(u) \rho_{env} u^{2} \over 8 \rho_{\rm p} r_{\rm p}}$, assuming, as in Grishin et al. (\citeyear{2019MNRAS.487.3324G}), that the relative velocity $u$ is a constant through the passage. This makes the drag coefficient also a constant through the passage and it can be taken out of the energy dissipation integral. As we discuss in Section \ref{capturerates}, this allows to solve for the critical impact parameter from which, following Grishin et al. (\citeyear{2019MNRAS.487.3324G}), we can calculate the capture rates.

\section{Derivation of $N_{\rm R_1<R<R_2}$ from $N_{\rm R\geqslant R_0}$}
\label{appendixC}
If we assume that the size distribution is given by
\begin{equation}
\begin{split}
n(r) = A_1 r^{-q_1}~{\rm if}~r_{min}<r<r_b\\
n(r) = A_2 r^{-q_2}~{\rm if}~r_b<r<r_{max},\\
\end{split}
\label{2s-sizedist2}
\end{equation}
with A$_2$ = A$_1$ r$_b^{q_2-q_1}$ (to have continuity at r$_b$), then, for R$_0 < $ r$_b$, 
\begin{equation}
\begin{split}
N_{\rm r \geqslant R_0} = 
A_1 \left[{1 \over 1-q_1}[{r_b^{1-q_1}}-R_0^{1-q_1}] + {r_b^{q_2-q_1}} {1 \over 1-q_2}[{r_{max}^{1-q_2}}-r_b^{1-q_2}]\right].
\end{split}
\label{NA1}
\end{equation}
Since we know $N_{\rm r \geqslant R_0}$, we can solve for A1 and use the power law above to get $N_{\rm R_1<R<R_2}$.\\

\begin{figure}
\begin{center}
\includegraphics[width=9cm]{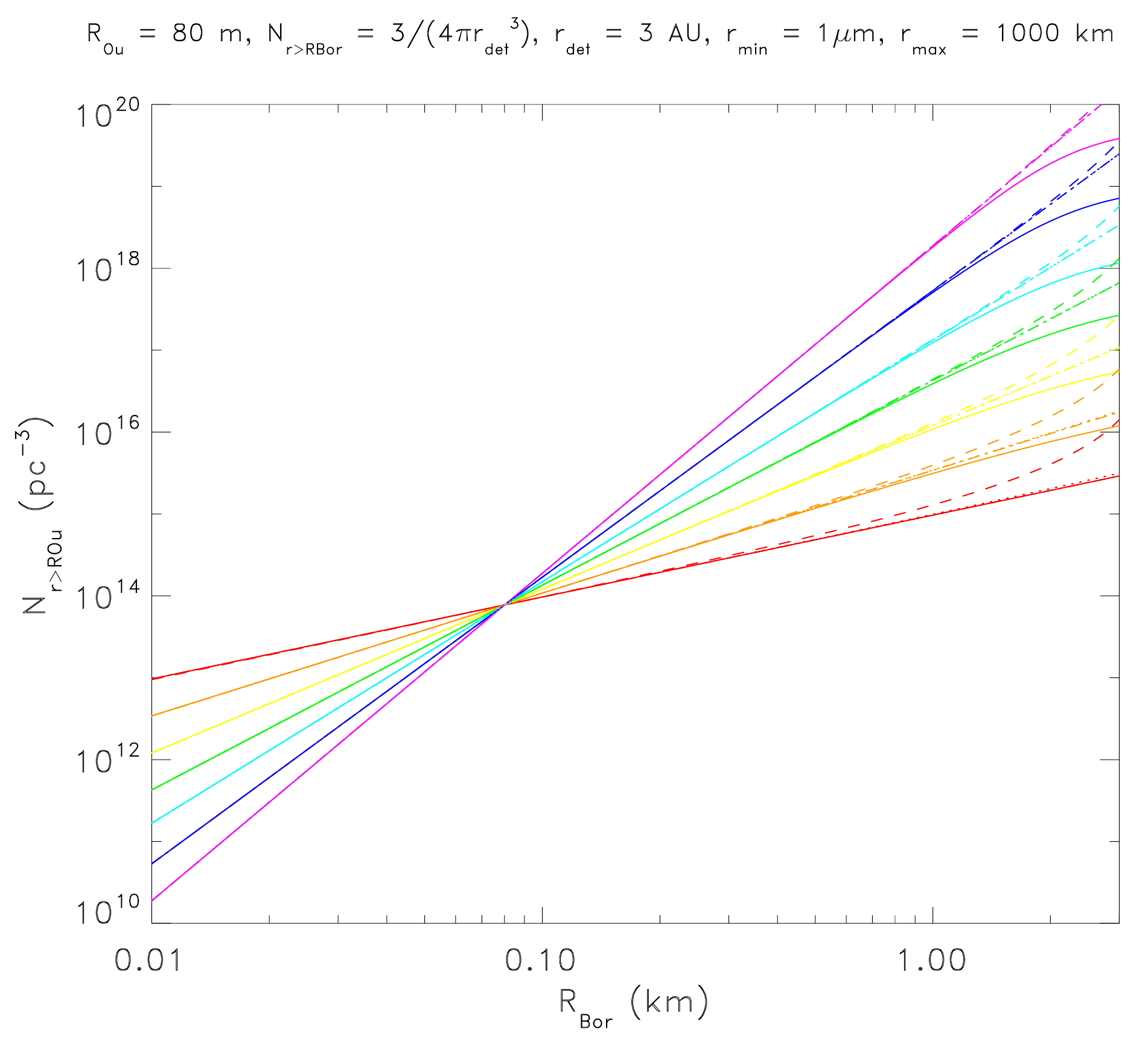}
\end{center}
\caption{Inferred cumulative number density of planetesimals with sizes equal or larger than 1I/'Oumuamua's, $N_{\rm R\geqslant R_{\rm Ou}}$ (adopting $R_{\rm Ou}$ = 80 m),  as a function of 2I/Borisov's radius, calculated using Equation \eqref{factor}. The different colors correspond to different values of the power-law index $q_{\rm 1}$, with $q_{\rm 1}$ = 2 (red), 2.5 (orange), 3 (yellow), 3.5 (green), 4 (light blue), 4.5 (dark blue), and 5 (pink). The line types correspond to the following power-law parameters:  $q_{\rm 2}$ = 2 and $r_{\rm b}$ = 3 km (solid),  $q_{\rm 2}$ = 6 and $r_{\rm b}$ = 3 km (dashed),  $q_{\rm 2}$ = 2 and $r_{\rm b}$ = 30 km (dotted), and $q_{\rm 2}$ = 6 and $r_{\rm b}$ = 30 km (dashed-dotted). These results are to be compared to the value of  $N_{\rm R\geqslant R_{\rm Ou}} = 2\cdot10^{15} {\rm pc}^{-3}$ adopted in Section \ref{NOumuamua}, based on the detection of 1I/'Oumuamua.}
\label{factorfig}
\end{figure}

The calculation in Section \ref{NOumuamua} adopts the cumulative number density of interstellar objects inferred from 1I/'Oumuamua's detection, estimated by Do et al. (\citeyear{2018ApJ...855L..10D}) to be $N_{\rm R\geqslant R_{\rm Ou}} = 2\cdot10^{15} {\rm pc}^{-3}$, for which we adopt $R_{\rm Ou} \sim$ 80 m. But 1I/'Oumuamua's brief visit left many open questions regarding its nature, like its shape, size, composition, and internal structure ('Oumuamua ISSI Team et al. \citeyear{2019NatAs...3..594O}, and references therein; Sekanina \citeyear{2019arXiv190108704S}; Moro-Mart{\'\i}n \citeyear{2019ApJ...872L..32M}) that result in uncertainties in its origin and the inferred number density distribution of interstellar objects (Moro-Mart{\'{\i}}n \citeyear {2018ApJ...866..131M}; \citeyear{2019AJ....157...86M}). 2I/Borisov, the second interstellar object that has been detected, has an unquestionable cometary nature, similar to the planetesimals expected to be ejected from protoplanetary disks (the majority of which will originate outside the snowline of their parent systems). We now assess to what degree the detection of 2I/Borisov can change the inferred number density of interstellar objects in the ISM and therefore the results listed above for $N_{\rm R_1<R<R_2}$. Following Jewitt \& Luu (\citeyear{2019arXiv191002547J}), the cumulative number density of interstellar objects similar or larger than 2I/Borisov would be $N_{\rm R\geqslant R_{\rm Bor}} = {1 \over V_{\rm det}}$, where $V_{\rm det} = {4 \over 3} \pi r_{\rm det}^3$ is the detection volume, with $r_{\rm det}$ = 3 AU, the distance from the Sun at which 2I/Borisov was discovered. Adopting the size distribution in Equation \eqref{2s-sizedist} and Equation (9) in Moro-Mart{\'\i}n (\citeyear{2019AJ....157...86M}), we get that the cumulative number density of interstellar objects similar or larger than 1I/'Oumuamua is  
\begin{equation}
\begin{split}
N_{\rm R\geqslant R_{\rm Ou}} = {3 \over 4 \pi r_{\rm det}^3} 
{{1 \over -q_1+1}\left[r_b^{-q_1+1}-R_{\rm Ou}^{-q_1+1}\right]+{r_b^{q_2-q_1} \over -q_2+1}\left[r_{max}^{-q_2+1}-r_b^{-q_2+1}\right] \over {1 \over -q_1+1}\left[r_b^{-q_1+1}-R_{\rm Bor}^{-q_1+1}\right]+{r_b^{q_2-q_1} \over -q_2+1}\left[r_{max}^{-q_2+1}-r_b^{-q_2+1}\right]}.
\end{split}
\label{factor}
\end{equation}
Because 2I/Borisov's cometary nature has made so far the determination of the size of the nucleus challenging, Figure \ref{factorfig} shows the results as a function of 2I/Borisov's size, adopting $R_{\rm Ou}$ = 80 m. Jewitt et al. (\citeyear{2020ApJ...888L..23J}) proposed that $R_{\rm Bor}$ = 0.2--0.5 km. If we were to adopt this range, we have that the resulting value for $N_{\rm R\geqslant R_{\rm Ou}}$ would be in the range (0.2--0.5)$\cdot10^{15} {\rm pc}^{-3}$ (for q$_{\rm 1}$ = 2) to (3--100)$\cdot10^{15} {\rm pc}^{-3}$ (for q$_{\rm 1}$ = 5), encompassing the adopted value of $N_{\rm R\geqslant R_{\rm Ou}} = 2\cdot10^{15} {\rm pc}^{-3}$ that was based on the detection of 1I/'Oumuamua alone (with the caveat that, given their very different properties, 1I/'Oumuamua and  2I/Borisov might not belong to the same population of interstellar comets).
  
\section{Derivation of $N_{\rm R_1<R<R_2}$ from m$_{\rm total}$}
\label{appendixD}
We can get the number density of planetesimals with radius R$_1<$ R $<R_2$  from the total mass density of planetesimals by first solving for the cumulative number density $N_{\rm r \geqslant R0}$ using Equation (17) of Moro-Mart\'{\i}n (\citeyear{2018ApJ...866..131M}), 
\begin{equation}
\begin{split}
{N_{\rm r \geqslant R_0} = m_{total}9(4\pi\rho)^{-1}}
{{1 \over 1-q_1}[{1 \over r_b^{3}}-{1 \over R_0^{3}}({R_0 \over r_b})^{4-q_1}]+{1 \over 1-q_2}[{1 \over r_{max}^{3}}({r_{max} \over r_b})^{4-q_2}-{1 \over r_b^{3}}] \over {3 \over 4-q_2}\left[\left({r_{max} \over r_{b}}\right)^{4-q_2}-1\right]+{3 \over 4-q_1}\left[1-\left({r_{min} \over r_{b}}\right)^{4-q_1}\right]},
\end{split}
\label{N-mtotal}
\end{equation}
assuming R$_0 < $ r$_b$, and then using the expressions derived in Appendix C to get $N_{\rm R_1<R<R_2}$ from $N_{\rm R\geqslant R_{\rm 0}}$.  \\

\end{document}